\documentclass[preprint2]{aastex61}

\PassOptionsToPackage{breaklinks=true}{hyperref}
\def\chandra    {\emph{Chandra}}
\def\xmm        {\emph{XMM}}
\def\vla        {VLA}
\def\mwa        {MWA}
\def\gmrt    {GMRT}

\def\rosat    {\emph{ROSAT}}
\def\asca    {\emph{ASCA}}

\def\as         {$^{\prime\prime}$}

\received{November 19, 2019}
\revised{January 8, 2020}
\accepted{\today}
\submitjournal{ApJ}
\shorttitle{A giant radio fossil in the Ophiuchus cluster}
\shortauthors{Giacintucci et al.}

\begin{document}

\title{DISCOVERY OF A GIANT RADIO FOSSIL IN THE OPHIUCHUS GALAXY CLUSTER}

\correspondingauthor{Simona Giacintucci}
\email{simona.giacintucci@nrl.navy.mil}

\author{S. Giacintucci}
\affiliation{Naval Research Laboratory, 
4555 Overlook Avenue SW, Code 7213, 
Washington, DC 20375, USA}
\author{M. Markevitch}
\affiliation{NASA/Goddard Space Flight Center,
Greenbelt, MD 20771, USA}
\author{M. Johnston-Hollitt}
\affiliation{International Centre for Radio Astronomy Research, Curtin University,
Bentley, WA 6102, Australia}
\affiliation{Peripety Scientific Ltd., PO Box 11355 Manners Street, Wellington 6142,
New Zealand}
\author{D. R. Wik}
\affiliation{Department of Physics and Astronomy, University of Utah, 115 South 1400 East, Salt Lake City, UT 84112, USA}
\author{Q. H. S. Wang}
\affiliation{Department of Physics and Astronomy, University of Utah, 115 South 1400 East, Salt Lake City, UT 84112, USA}
\author{T. E. Clarke}
\affiliation{Naval Research Laboratory, 
4555 Overlook Avenue SW, Code 7213, 
Washington, DC 20375, USA}
\begin{abstract}
The Ophiuchus galaxy cluster exhibits a curious concave gas density
discontinuity at the edge of its cool core. It was discovered in the {\em Chandra}
X-ray image by Werner and collaborators, who considered a possibility of it
being a boundary of an AGN-inflated bubble located outside the core, but
discounted this possibility because it required much too powerful an AGN
outburst. Using low-frequency (72--240 MHz) radio data from \mwa/GLEAM and
\gmrt, we found that the X-ray structure is, in fact, a giant cavity in the
X-ray gas filled with diffuse radio emission with an extraordinarily steep
radio spectrum. It thus appears to be a very aged fossil of the most
powerful AGN outburst seen in any galaxy cluster ($pV\sim 5\times 10^{61}$
erg for this cavity). There is no apparent diametrically opposite counterpart
either in X-ray or in the radio. It may have aged out of the observable
radio band because of the cluster asymmetry. At present, the central AGN
exhibits only a weak radio source, so it should have been much more powerful
in the past to have produced such a bubble. The AGN is currently starved of
accreting cool gas because the gas density peak is displaced by core
sloshing. The sloshing itself could have been set off by this extraordinary
explosion if it had occurred in an asymmetric gas core. This {\em
  dinosaur}\/ may be an early example of a new class of sources to be
uncovered by low-frequency surveys of galaxy clusters.

\end{abstract}


%===================================================================

\section{Introduction}

The hot and massive Ophiuchus cluster of galaxies, the second X-ray
brightest cluster in the sky, has a cool core with a very steep temperature
drop from 10 keV to below 1 keV (e.g., P\'erez-Torres et al.\ 2009,
hereafter P09; Werner et al.\ 2016, hereafter W16). A high-resolution
  X-ray image of the Ophiuchus core region, with interesting features
  marked, is shown in Fig.\ 1, derived using the same data as in W16. The
core exhibits several prominent sloshing X-ray cold fronts -- concentric gas
density edges at various radii (Ascasibar \& Markevitch 2006; ZuHone et al.\ 
2010; Million et al.\ 2010; W16). The core also exhibits a radio synchrotron
minihalo (Govoni et al.\ 2009, Murgia et al.\ 2010, hereafter G09 and M10),
similarly to most cool cores in massive clusters (Giacintucci et al.\ 2017).
The same population of ultra-relativistic electrons may explain the reported
nonthermal hard X-ray emission (Eckert et al.\ 2008; Nevalainen et al.\ 
2009; Fujita et al. 2008). The {\em Chandra}\/ high-resolution X-ray image
revealed that sloshing has displaced the peak of the cluster cool gas away
from the cD galaxy (ZuHone et al.\ 2010; Million et al.\ 2010), apparently
depriving the currently active galactic nucleus (AGN) of its accretion fuel,
which is probably why the AGN is currently weak both in the radio and in the
X-ray (Hamer et al.\ 2012; W16).

%%%%%%%%%%%
%%%%%%%%%%%%%%%%%%%%%%%%%%%%%%%%%%%%%%%%%%%%%%%%%%%%%%%%%%%%%%%%%%%%%
\begin{figure}
\centering \epsscale{1}
\includegraphics[width=8.5cm]{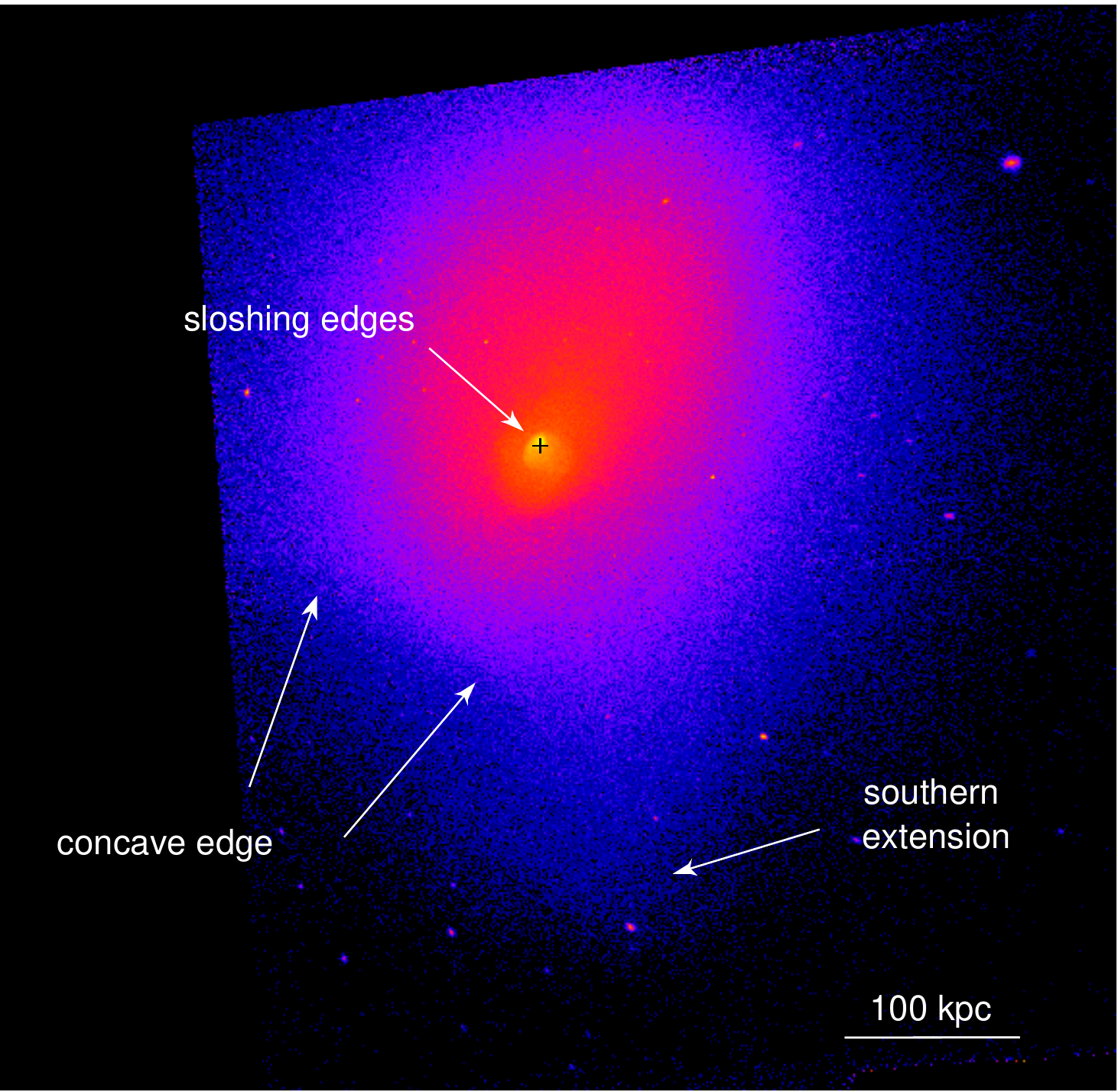} 
\caption{{\em Chandra} X-ray image of the Ophiuchus central region in the 0.5--4 keV band, 
with main features marked. Black cross marks the cD galaxy; the X-ray peak is displaced 
from that position, apparently by vigorous gas sloshing, as evidenced by several 
prominent concentric brightness edges at different radii. The subtle concave edge, 
noticed by W16, is the subject of this work.}
%Fig. 1---\chandra\ X-ray image of the Ophiuchus central region in the 
%0.5--4 keV band. We used the same data as in W16. The cross marks the
%  position of a cD galaxy; the X-ray peak is displaced from that position,
%  apparently by gas sloshing, as evidenced by several prominent concentric
%  brightness edges at different radii (ZuHone et al.\ 2010; Million et al.\ 
%  2010; W16). The concave edge, noticed by W16, is the subject of this work.}
\label{fig:images_chandra}
\end{figure}
%%%%%%%%%%%%%%%%%%%%%%%%%%%%%%%%%%%%%%%%%%%%%%%%%%%%%%%%%%%%%%%%%%%%%

%%%%%%%%%%

Another curious feature noted by W16 in their deep {\em Chandra} X-ray image
of the Ophiuchus center is a concave gas density edge at the boundary of the
cool core (Fig.\ 1). W16 briefly considered a possibility that this is a
bubble blown by jets from the central AGN, similar to Perseus (Fabian et al.
2003), Hydra A (Nulsen et al.\ 2005), MS\,0735+74 (McNamara et al.\ 2005)
and numerous other examples found in clusters and massive elliptical
galaxies (e.g., Rafferty et al.\ 2006).
% and Centaurus (Sanders et al.\ 2016). 
Such X-ray cavities are usually filled with extended radio synchrotron
emission from the ultrarelativistic electrons ejected by the AGN.  The radio
spectrum of this emission steepens as these electrons cool on 10--100 Myr
timescales (e.g., Murgia et al. 2011). M10 did find a hint of such extended
radio emission near the edge of the putative cavity (their ``source E''),
but interpreted it as a steep-spectrum outlying region of the central radio
minihalo.  W16 estimated the diameter of the cavity from the curvature of
the X-ray edge and the $pV$ work\footnote{$p$ is the pressure of the gas
  surrounding the cavity and $V$ is the cavity's volume.} required to
displace the intracluster medium (ICM) to create such a cavity, $5\times
10^{61}$ erg, which turns out to be several times greater than that for the
most powerful unambiguous AGN outburst known, MS\,0735+74 (McNamara et al.\ 
2005). They argued that such an explosion would very likely have wiped out
the cool core, which still exists and exhibits sharp gradients of entropy
and metallicity.  Therefore, they concluded that the feature is more
plausibly a hydrodynamic effect of a cluster merger, not unlike the
Kelvin-Helmholtz eddy at the edge of the Perseus core (Walker et al.\ 2017),
which are occasionally seen in hydro simulations of cluster core sloshing
(e.g., Ascasibar \& Markevitch 2006).

In this paper, we inspect low-frequency radio images from the Murchison 
Widefield Array (MWA), Giant Metrewave Radio Telescope (GMRT) and Very 
Large Array (VLA) --- where the aged relativistic particles from past AGN activity
might still leave trace emission --- compare them with the X-ray images, and uncover
a  
%unmistakeable 
fossil of an extraordinarily powerful AGN outburst.

We use a $\Lambda$CDM cosmology with H$_0$=70 km s$^{-1}$ Mpc$^{-1}$,
$\Omega_m=0.3$ and $\Omega_{\Lambda}=0.7$. At the redshift of 
Ophiuchus ($z=0.028$), $1^{\prime\prime}$ = 0.562 kpc. 
All errors are quoted at the $68\%$ confidence level. The radio spectral 
index $\alpha$ is defined according to $S_{\nu} \propto \nu^{-\alpha}$, 
where $S_{\nu}$ is the flux density at the frequency $\nu$. 

\section{RADIO DATA}
\label{sec:radio}

Intrigued by the W16 estimate of the enormous energy output required to
displace the cluster X-ray gas if the concave X-ray edge were a cavity, we
have searched the low-frequency radio data for faint diffuse signal from
aged ultrarelativistic particles ejected by this hypothetical AGN explosion.
Indeed, the low-frequency images from the MWA GLEAM survey% 
\footnote{MWA GaLactic and Extragalactic All-sky
  MWA (Wayth et al. 2015, Hurley-Walker et al. 2017).}
in several bands covering 72--230 MHz reveal a large, extended radio source
%with a very steep spectrum 
cospatial with the putative X-ray cavity. This
source was not detected in previously published radio images (P09, G09, M10)
and extends far beyond the source E in M10. This prompted us to reanalyze
and combine the archival \gmrt\ data used in P09 and M10 in order to reach higher
sensitivity and investigate the nature of this emission and its connection
to the X-ray cavity. Below we compare our rederived, lower-noise GMRT images
at 153 MHz and 240 MHz to the GLEAM images, as well as to a 74 MHz image
from the VLSSr survey.% 
\footnote{\vla\ Low-Frequency Sky Survey Redux (Lane et al.\ 2014).}
We also use an image at 1477 MHz from the archival \vla\ D-configuration
observation that was rederived in Giacintucci et al.\ (2019, hereafter G19) 
from the data used in G09 and M10.

The observations used in our analysis provide sensitivity to angular structures 
as large as $36^{\prime}$ (VLSSr) or larger ($44^{\prime}-68^{\prime}$ 
for the GMRT and several degrees for GLEAM), which correspond to a physical scale 
of at least 1.2 Mpc at the redshift of Ophiuchus. The largest linear scale 
detectable by the VLA D-configuration observation at 1477 MHz is significantly 
smaller ($\sim 500$ kpc).

\subsection{\gmrt}

We obtained from the GMRT archive the observations of the Ophiuchus cluster
at 153 MHz and 240 MHz used by M10 and P09 and reduced them using the
NRAO\footnote{National Radio Astronomy Observatory.} Astronomical Image
Processing System (AIPS). Details on these observations are summarized in
Table 1.

The data were collected in spectral-line observing mode, using the GMRT
hardware backend, with a total observing bandwidth of 8 MHz. We carefully
inspected the data and found that all observations were partially impacted
by radio frequency interference (RFI). We used RFLAG to excise RFI-affected
visibilities, followed by manual flagging to remove residual bad data. Gain
and bandpass calibrations were applied using the standard primary
calibrators 3C286 at 153 MHz and 3C286 and 3C48 at 240 MHz. The Scaife \&
Heald (2012) scale was adopted in SETJY to set their flux densities at these
frequencies.  Phase calibrators, observed several times during the
observation, were used to calibrate the data in phase. To reduce the size of
the data sets, minimizing at the same time the effects of bandwidth
smearing, the calibrated visibilities were averaged appropriately in
frequency, with 14 channels at 240 MHz and 21 channels at 153 MHz, each
0.375 MHz and 0.25 MHz wide, respectively. A number of phase
self-calibration iterations, followed by a final self-calibration step in
amplitude and phase, were applied to the target visibilities. During the
self-calibration process, images were made using wide-field imaging,
decomposing the primary beam area into a large number of smaller facets.
Additional facets were placed on outlier bright sources out to a distance of
10 degrees from the phase center. The final images were produced using
wide-field and multi-resolution imaging in AIPS with the Briggs robust
weighting (Briggs 1995) set to 0. At 240 MHz, we combined the two final
self-calibrated data sets in the $uv$ plane to produce the final images. The
root mean square ({\em rms}) noise levels reached in our full-resolution
images are reported in Table 1. Our values are significantly lower ($\sim
40\%-50\%$) than the noise levels reached by M10 and P09.  
Finally, correction for the \gmrt\ primary beam response was applied using PBCOR in
AIPS\footnote{http://www.ncra.tifr.res.in:8081/{\textasciitilde}ngk/
\\
primarybeam/beam.html.}.

Residual amplitude errors, $\sigma_{\rm amp}$, are estimated to be within
$15\%$ at 153 MHz and $10\%$ at 240 MHz (Chandra, Ray \& Bhatnagar 2004). 
Errors on measured flux densities were calculated as 
\begin{equation}
\sigma_{S_{\nu}} = \sqrt{(\sigma_{\rm amp} \times S_{\nu})^2 + (\sigma_{rms} \times \sqrt{N_{\rm beam}})^2},
\end{equation}
\label{eq:err}
where $S_{\nu}$ is the source flux density at the frequency $\nu$,
$\sigma_{rms}$ is the noise level in the image, and $N_{\rm beam}$ is the
number of beams crossing the source.

%%%%%%%%%%%%%%%%%%%%%%%%%%%%%%%%%%%%%%%%%%%%%%%%%%%%%%%%%%%%%%%%%%%%%
\begin{deluxetable*}{cccccccccc}
\tablecaption{Newly analyzed GMRT observations}
\label{tab:radioobs}
\tablehead{
\colhead{Project} &  \colhead{Frequency} & \colhead{Bandwidth}& \colhead{Observation} & \colhead{Time}  & \colhead{FWHM} & \colhead{{\em rms}} &  \colhead{$\Theta_{\rm LAS}$}\\
\colhead{code}    & \colhead{(MHz)}     & \colhead{(MHz)}    &  \colhead{date}    & \colhead{(hour)}& \colhead{($^{\prime \prime} \times^{\prime \prime}$)} &  \colhead{(mJy)}  & \colhead{($^{\prime}$)} \\
}
\startdata
14MPA03       & 153   & 8  & 2008 Aug 21               &  3.8  & $27\times22$ & 3.0   & 68  \\
RAS + 14MPA03 & 240   & 8  & 2008 May 24, Aug 23,28    & 11.9  & $15\times12$ & 0.55  & 44  \\
%VLA--D  & AC261         & 1477  & 25 & 1990 Jan 25               &  0.5  & $92\times39$ & 0.1   & 15  \\
\enddata \tablecomments{Column 1: project code. Columns 2 and 3: Observing
  central frequency and bandwidth. Column 4: Observation dates. Column 5:
  Total time on source. Columns 6 and 7: Full width half maximum (FWHM) of
  the radio beam and {\em rms} level ($1\sigma$) of the final images at full
  array resolution. Column 8: largest angular scale detectable by the array.
  The cluster center coincides with the radio point source embedded in the
  diffuse minihalo and associated with the central cD galaxy (yellow cross in
Fig.~\ref{fig:images1}{\em a}).}
\end{deluxetable*}
%%%%%%%%%%%%%%%%%%%%%%%%%%%%%%%%%%%%%%%%%%%%%%%%%%%%%%%%%%%%%%%%%%%%%

\subsection{\gmrt, \mwa\ and \vla\ images}
\label{sec:radioim}

%%%%%%%%%%%%%%%%%%%%%%%%%%%%%%%%%%%%%%%%%%%%%%%%%%%%%%%%%%%%%%%%%%%%%
\begin{figure*}
\centering \epsscale{1.1}
\includegraphics[width=9cm]{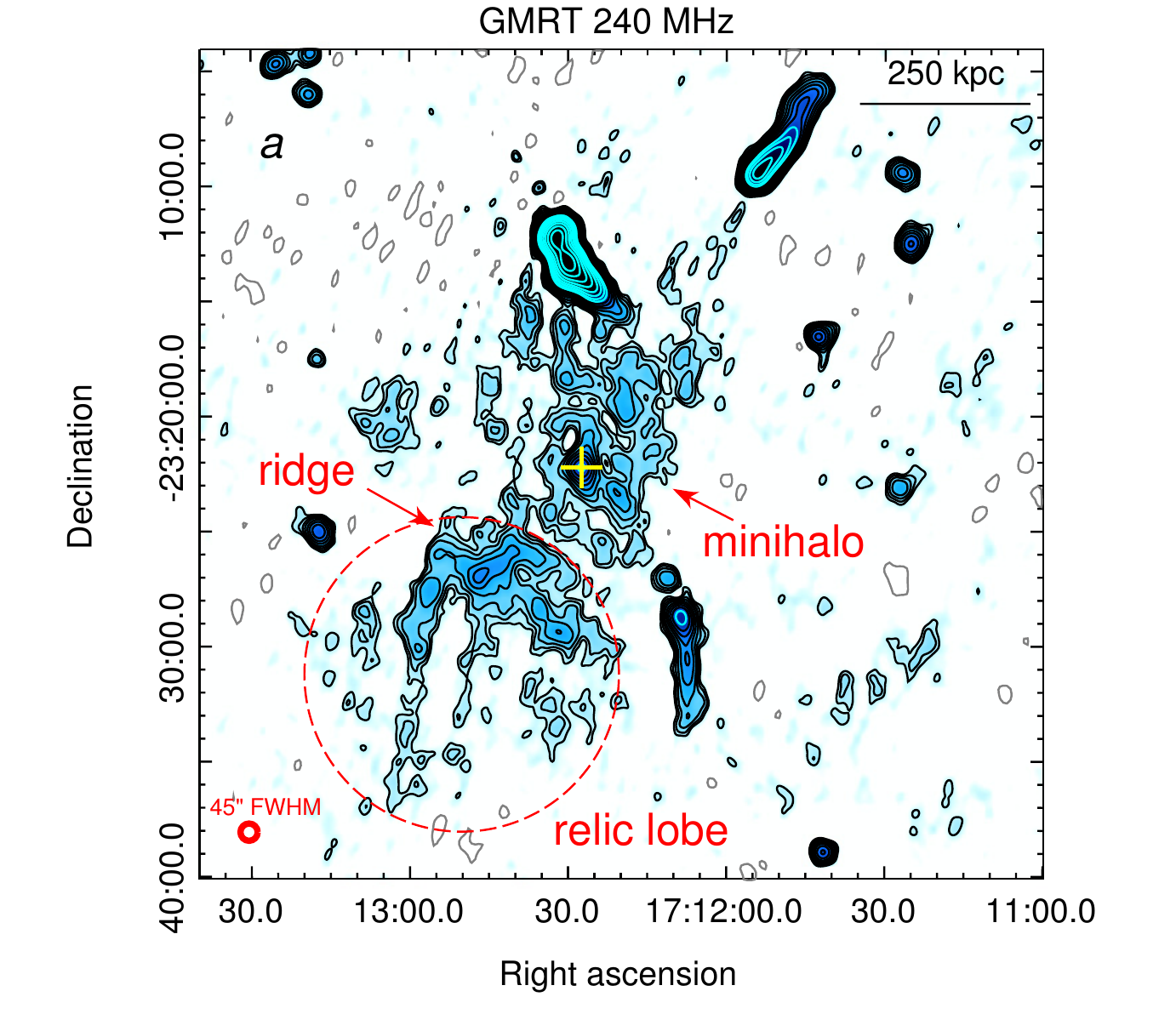} \hspace{-1cm}
\includegraphics[width=9cm]{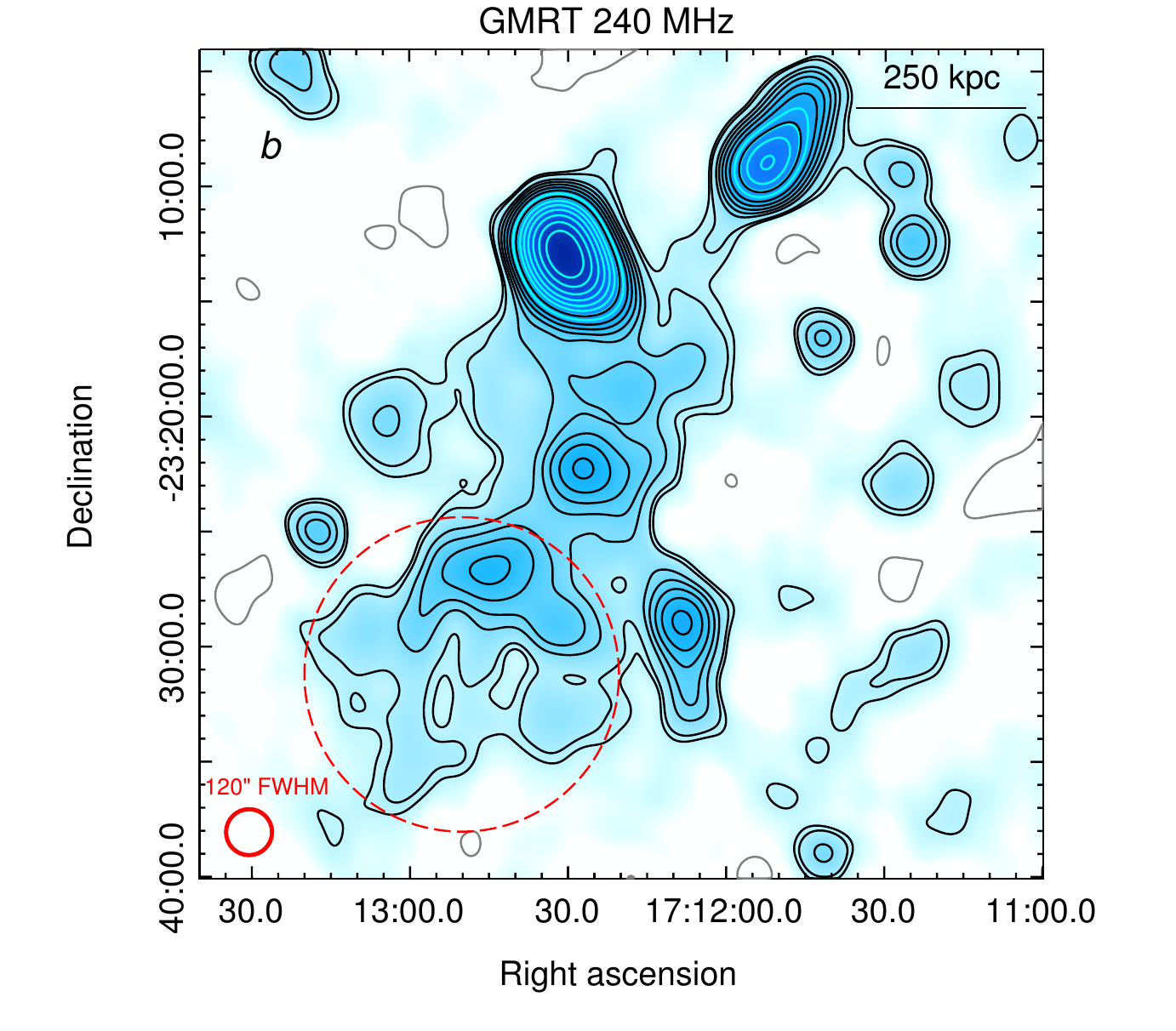} \smallskip
\caption{GMRT radio images at 240 MHz at the resolution of
  $45^{\prime\prime}$ ({\em a}) and $120^{\prime\prime}$ ({\em b}). The restoring beam
  is also shown in red in the bottom-left corner of each image.  The {\em
    rms} noise $1\sigma$ level is 1.3 beam$^{-1}$ and 3.3 mJy
  beam$^{-1}$, respectively. Contours (black and cyan) are spaced by a
  factor of $\sqrt{2}$, starting from $+3\sigma$.  Contours at $-3\sigma$
  are shown as grey. The radius of the dashed, red circle is
  $6^{\prime\prime}.8=230$ kpc. The yellow cross in panel {\em a}
marks the position of the central cD galaxy.}
\label{fig:images1}
\end{figure*}
%%%%%%%%%%%%%%%%%%%%%%%%%%%%%%%%%%%%%%%%%%%%%%%%%%%%%%%%%%%%%%%%%%%%%

%%%%%%%%%%%%%%%%%%%%%%%%%%%%%%%%%%%%%%%%%%%%%%%%%%%%%%%%%%%%%%%%%%%%%
\begin{figure}
\centering
\epsscale{1.1}
\includegraphics[width=9cm]{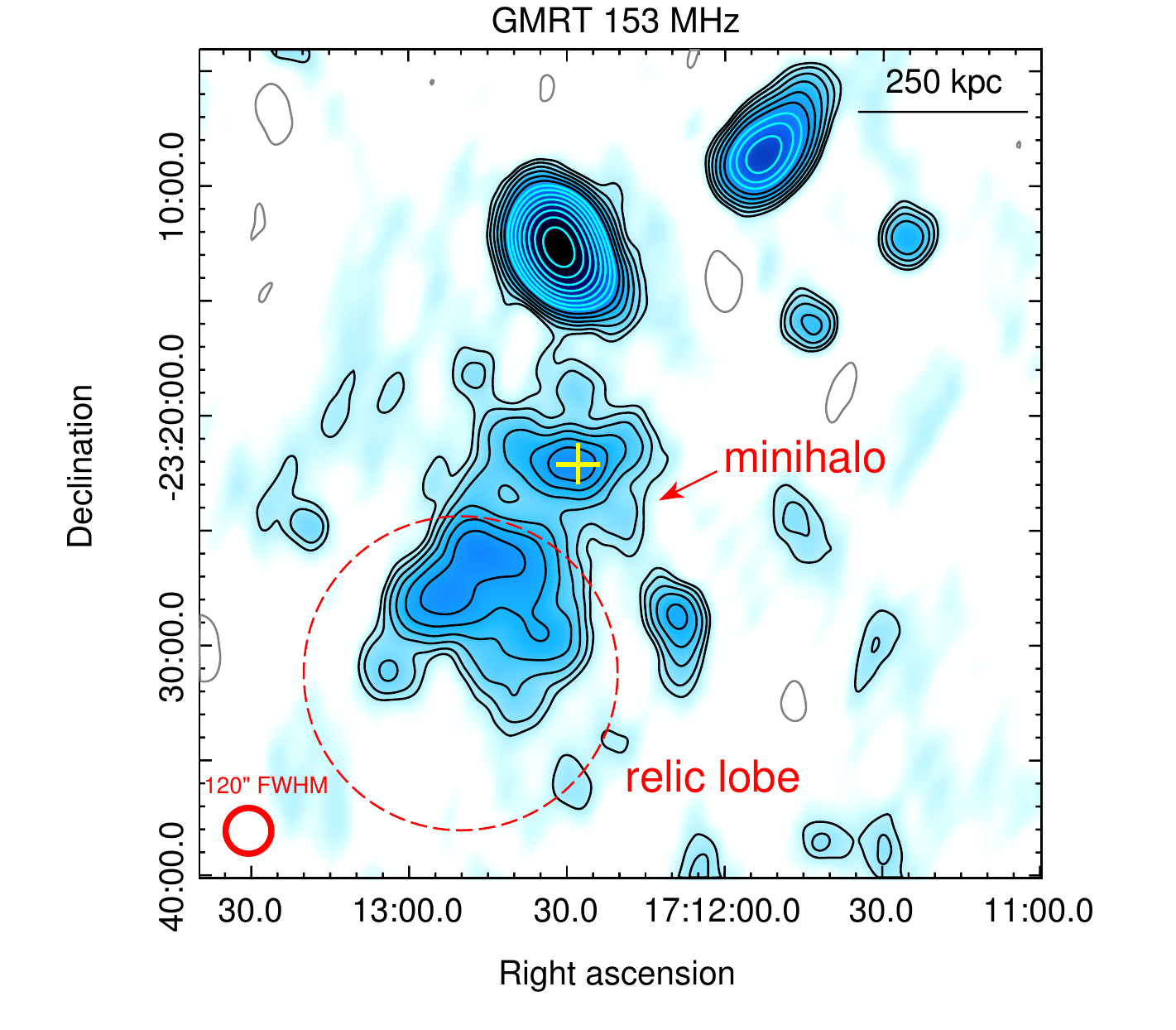}
%\smallskip
\caption{GMRT radio image at 153 MHz at $120^{\prime\prime}$ resolution.
  The restoring beam is also shown in red in the bottom-left corner. The
  {\em rms} noise $1\sigma$  level is 12 mJy beam$^{-1}$. Contours are spaced
  by a factor of $\sqrt2$, starting from $+3\sigma$. Contours at $-3\sigma$
  are shown as grey.  The radius of the dashed, red circle is
  $6^{\prime\prime}.8=230$ kpc (as in Fig.~\ref{fig:images1}). 
The yellow cross marks the position of the central cD galaxy.}
\label{fig:images2}
\end{figure}
%%%%%%%%%%%%%%%%%%%%%%%%%%%%%%%%%%%%%%%%%%%%%%%%%%%%%%%%%%%%%%%%%%%%%

%%%%%%%%%%%%%%%%%%%%%%%%%%%%%%%%%%%%%%%%%%%%%%%%%%%%%%%%%%%%%%%%%%%%%
\begin{figure*}
\centering
\epsscale{1.1}
\includegraphics[width=8.5cm]{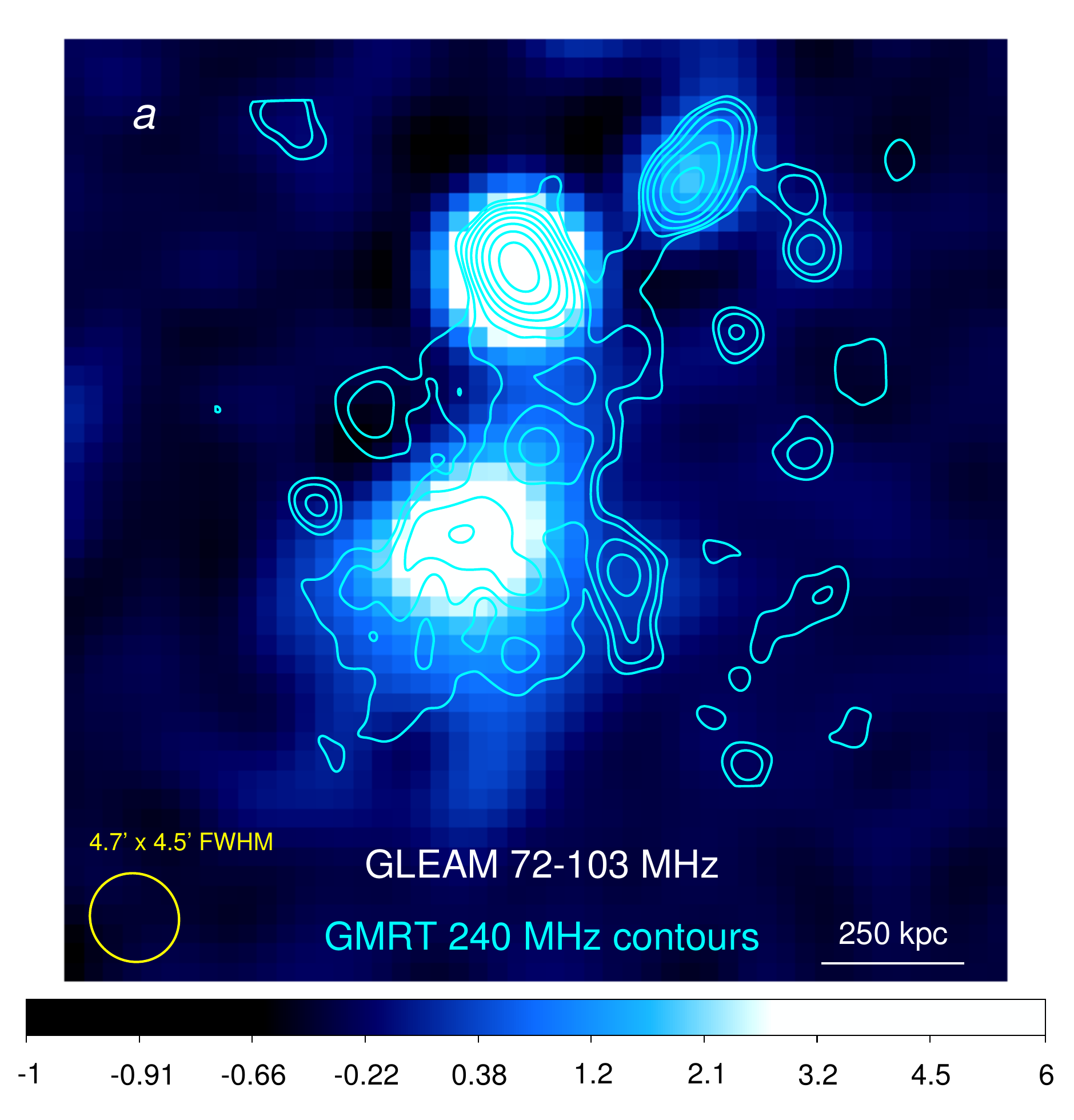}
\includegraphics[width=8.5cm]{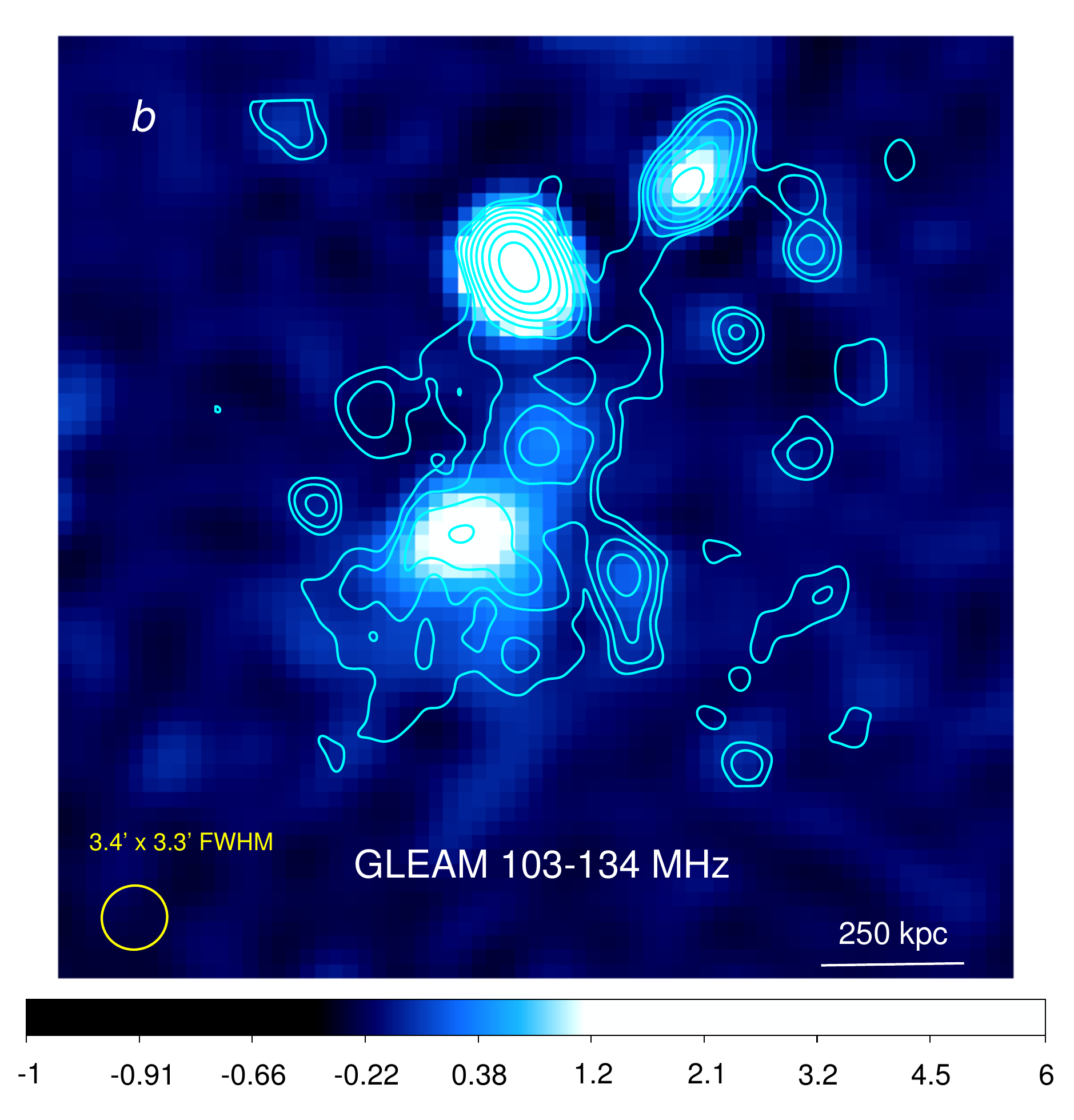}
\includegraphics[width=8.5cm]{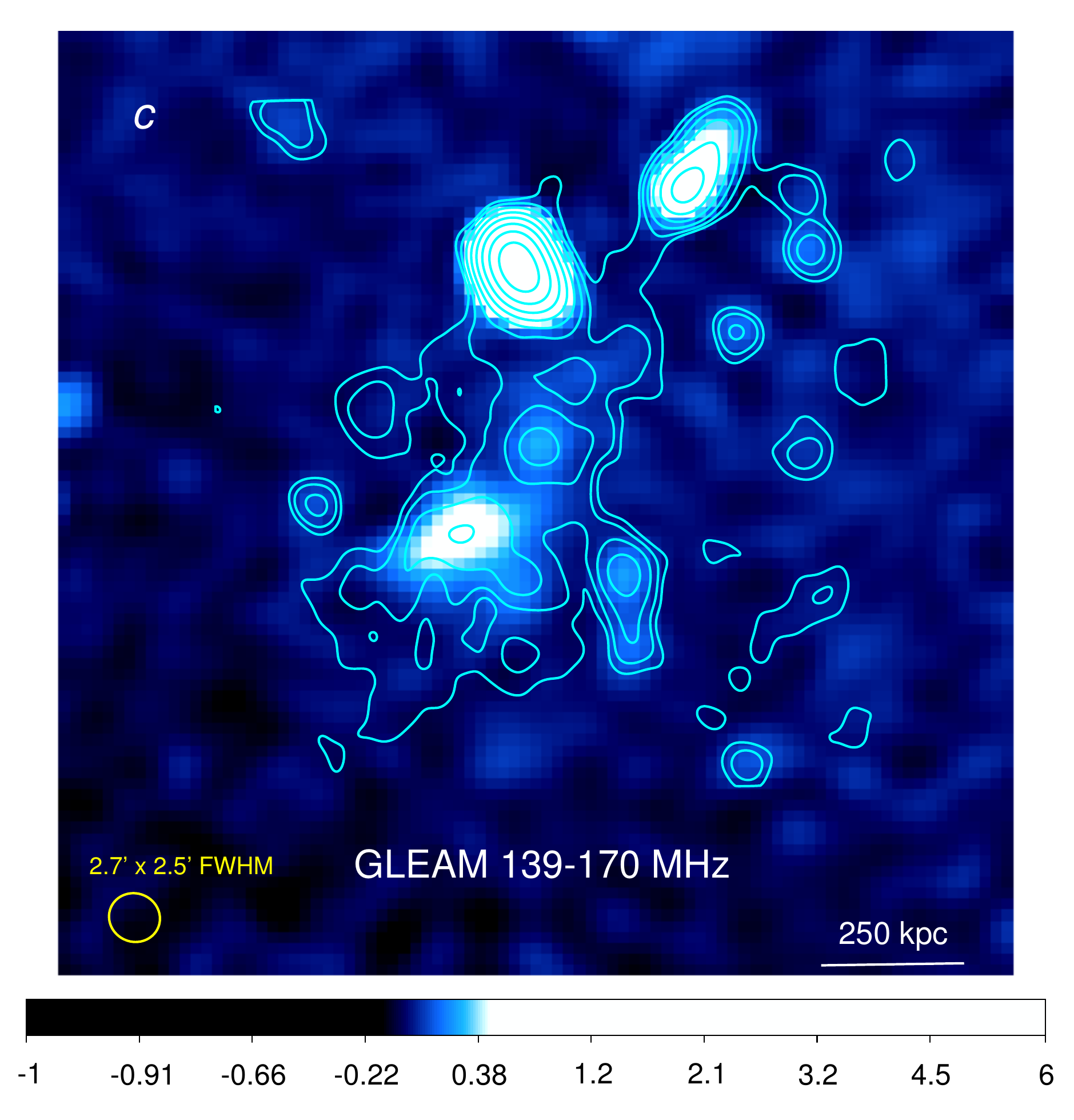}
\includegraphics[width=8.5cm]{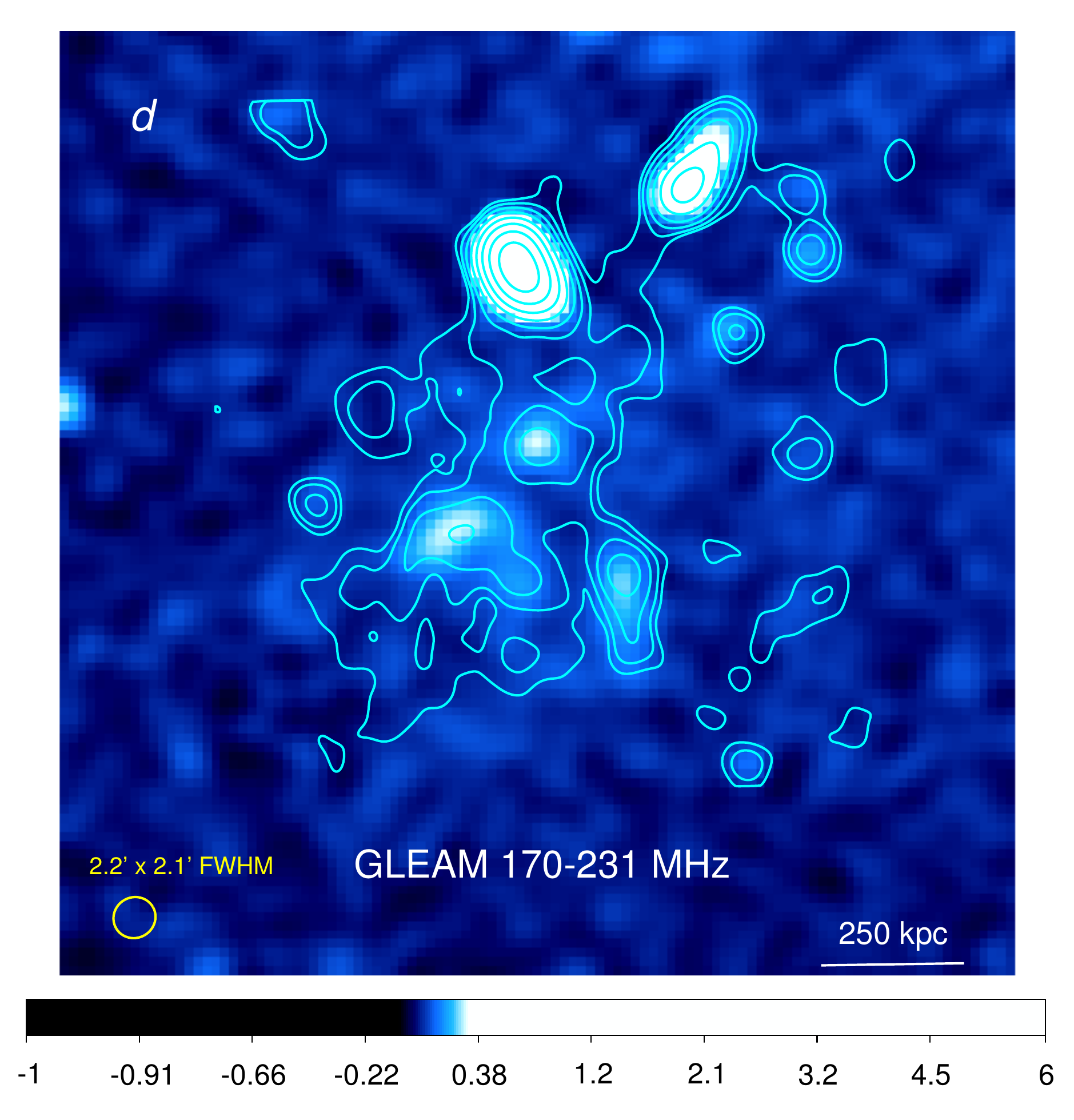}
\smallskip
\caption{GLEAM images at the central frequencies of (a) 88 MHz,
  (b) 118 MHz, (c) 154 MHz, (d) 200 MHz are shown by color; color scale
  units are Jy. Contours overlay the GMRT\, 240 MHz, $2'$ resolution
  image from Fig.~\ref{fig:images1}{\em b}. The contours are 
spaced by a factor 2, starting from $+3\sigma$. In each panel, 
the beam size of the GLEAM image is shown by a yellow ellipse.}
\label{fig:gleam}
\end{figure*}
%%%%%%%%%%%%%%%%%%%%%%%%%%%%%%%%%%%%%%%%%%%%%%%%%%%%%%%%%%%%%%%%%%%%%

Figure \ref{fig:images1} presents our GMRT images at 240 MHz at a
resolution of $45^{\prime\prime}$ ({\em a}) and $120^{\prime\prime}$ ({\em b}).  
The radio emission is dominated by three bright extended radio
galaxies. A detailed morphological and spectral study of these sources has
been presented by M10. The central diffuse minihalo (G09, M10) 
is well detected in our images, where it occupies an area of $\sim 200$
kpc in radius. The location of the central cD galaxy (yellow cross)
coincides with the faint radio point source embedded in the minihalo 
emission (M10, W16).

A distinct region of extended emission (hereafter referred to as relic
lobe) is visible South-East of the cluster center. Its brightest part
(ridge) coincides with source E in M10, but we see a considerably
larger source than in M10, thanks to the higher sensitivity of our GMRT 
images. The relic lobe extends out to a distance $d\sim17^{\prime}$ 
($\sim 570$ kpc) from the cluster center and has a radius of $r\sim 
6.^{\prime}8$ ($\sim 230$ kpc). A GMRT\/ image at 153 MHz at the
resolution of $120^{\prime\prime}$ is shown in Figure \ref{fig:images2}. 
Due to the lower sensitivity of this image, only the innermost part of the 
minihalo ($r\sim 130$ kpc) and the brightest portion of the relic lobe 
($d\sim 400$ kpc) are well detected.

\begin{figure}
\centering
\epsscale{1.1}
\includegraphics[width=9cm]{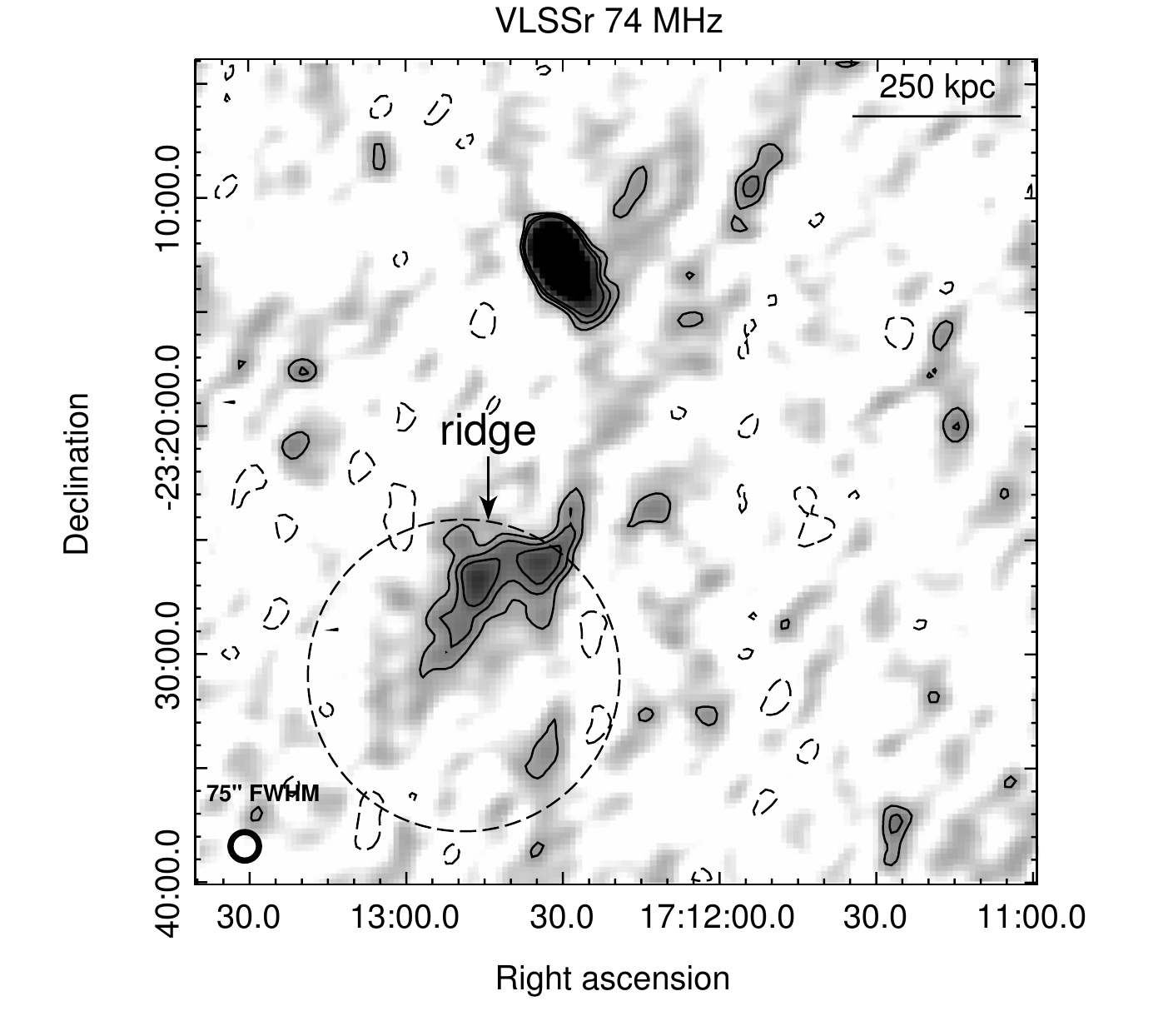}
\caption{Image at 74 MHz from the VLSSr (gray scale and contours). 
The restoring beam is $75^{\prime\prime}$ (also shown in black in the bottom-left corner) 
and the {\em rms} noise $1\sigma$ level is 180 mJy beam$^{-1}$. Contours
are $-2\sigma$ (dashed), $+2\sigma$, $+3\sigma$ and $+4\sigma$. The size of
the dashed circle is as in Fig.~\ref{fig:images1}.}
\label{fig:vlssr}
\end{figure}

\begin{figure}
\centering
\epsscale{1.1}
\includegraphics[width=7.5cm]{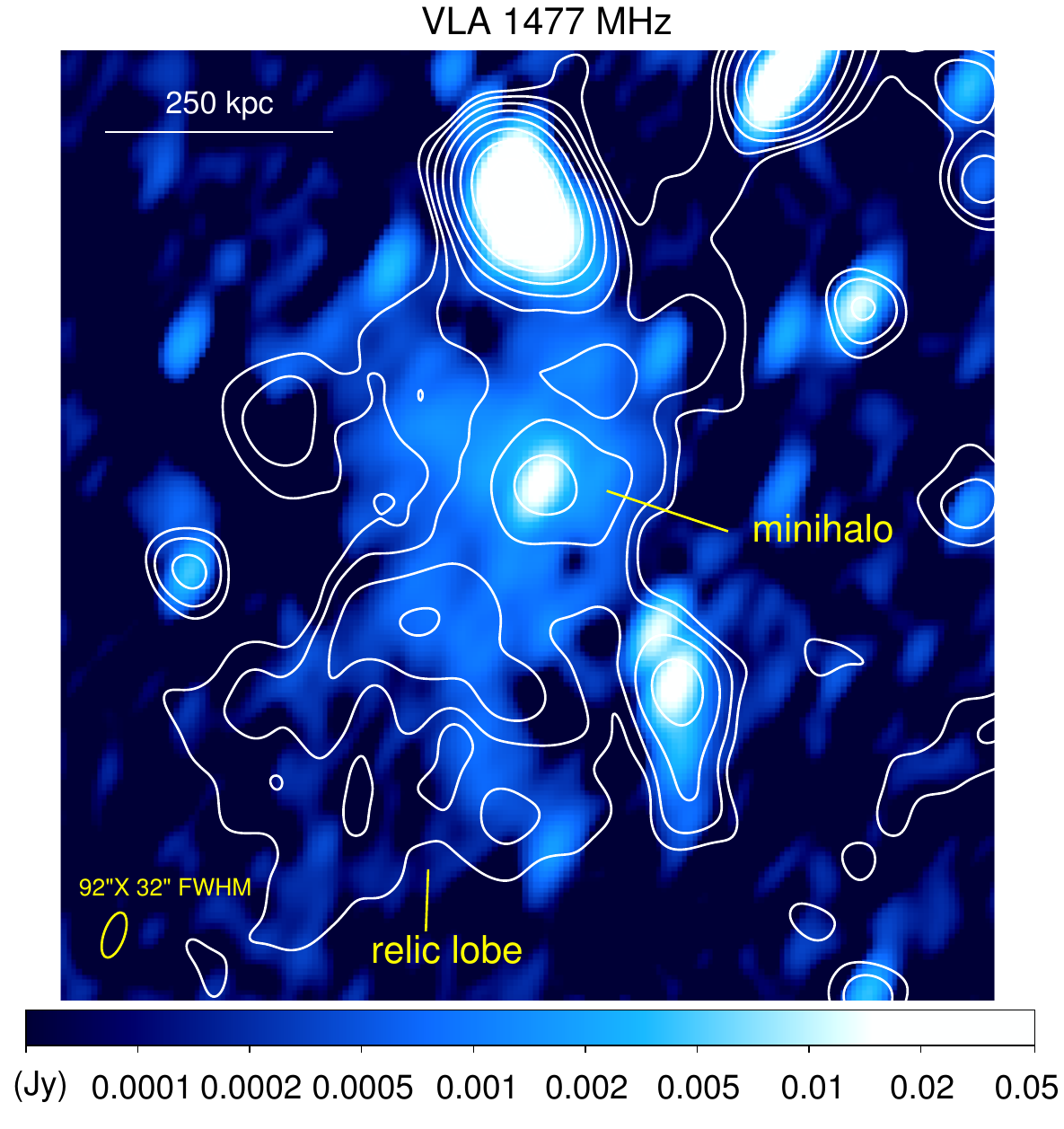}
\caption{GMRT\, 240 MHz contour image at $2^{\prime}$ resolution, 
overlaid on the VLA image at 1477 MHz. The restoring beam of the VLA 
image is $92^{\prime\prime}\times32^{\prime\prime}$, 
in p.a. $-25^{\circ}$ (also shown in yellow) and the $1\sigma$ noise 
is 0.1 mJy beam$^{-1}$.}
\label{fig:vla}
\end{figure}

In Figure \ref{fig:gleam}, we overlay our GMRT 240 MHz contours on the
low-frequency continuum images from GLEAM in the 72-103 MHz, 103-134 MHz, 
139-170 MHz and 170-231 MHz sub-bands. 
At the lowest GLEAM frequencies 
(panels {\em a} and {\em b}), the relic lobe is fully detected with 
good spatial coincidence with the GMRT image. The 139-170 MHz 
and 170-231 MHz sub-band images, instead, detect well only its northern 
ridge. The ridge is also visible in the VLSSr image at 74 MHz, shown 
in Fig.~\ref{fig:vlssr}.  Its peak emission is detected at a 
$4\sigma$ level ($1\sigma=180$ mJy beam$^{-1}$). The fainter 
emission from the minihalo is not detected at this sensitivity level. 
Finally, in Figure \ref{fig:vla} we compare the 240 MHz emission 
to a higher-frequency VLA image at 1477 MHz (from G19),
showing the central minihalo and only hints of the much lower 
surface brightness emission from the relic lobe.

%%%%%%%%%%%%%%%%%%%%%%%%%%%%%%%%%%%%%%%%%%%%%%%%%%%%%%%%%%%%%%%%%%%%%
%%%%%%%%%%%%%%%%%%%%%%%%%%%%%%%%%%%%%%%%%%%%%%%%%%%%%%%%%%%%%%%%%%%%%

%%%%%%%%%%%%%%%%%%%%%%%%%%%%%%%%%%%%%%%%%%%%%%%%%%%%%%%%%%%%%%%%%%%%%
%%%%%%%%%%%%%%%%%%%%%%%%%%%%%%%%%%%%%%%%%%%%%%%%%%%%%%%%%%%%%%%%%%%%%

%%%%%%%%%%%%%%%%%%%%%%%%%%%%%%%%%%%%%%%%%%%%%%%%%%%%%%%%%%%%%%%%%%%%%
\begin{figure}
\centering
\epsscale{1.1}
\includegraphics[width=7.5cm]{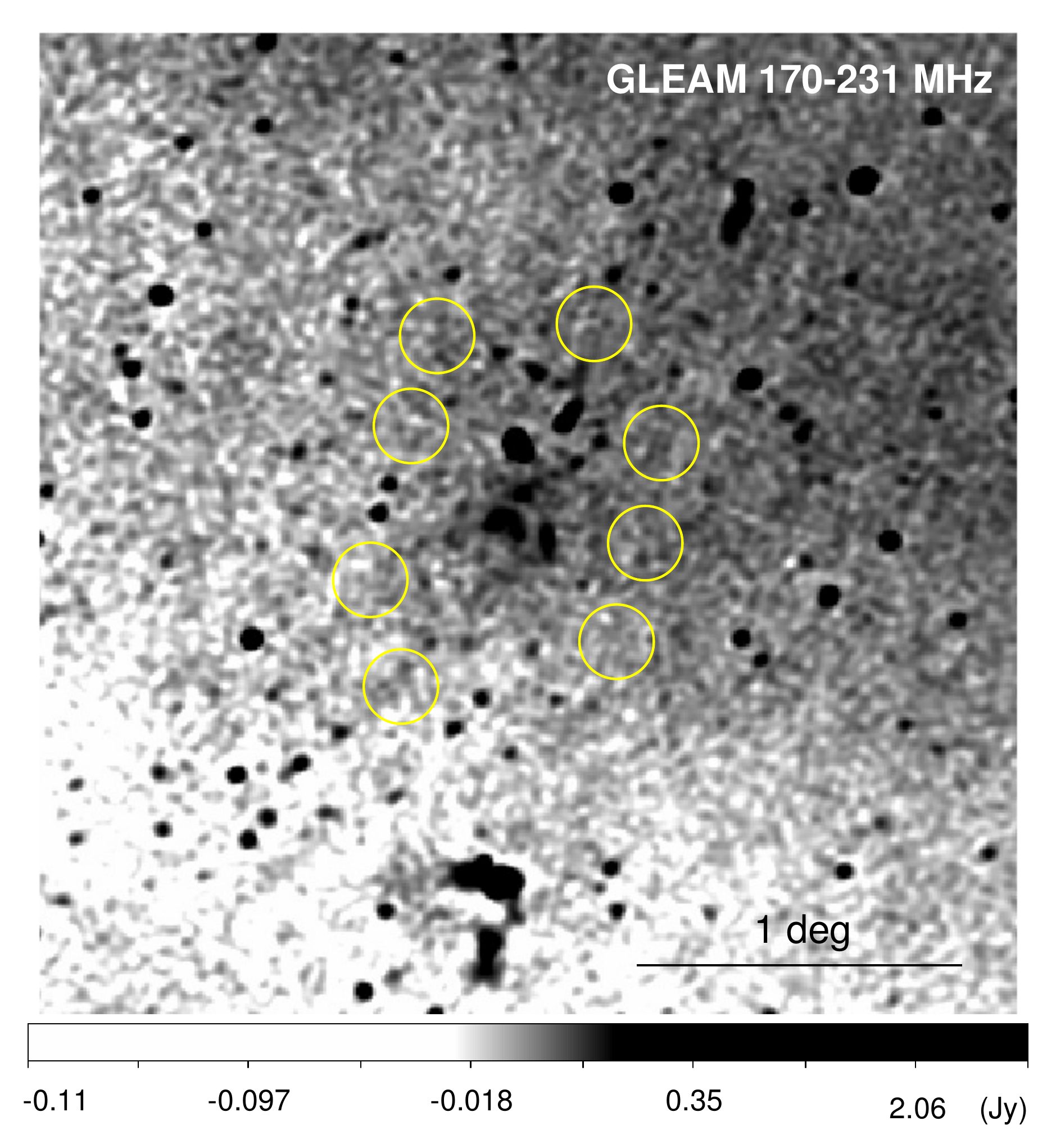}
\caption{GLEAM $3^{\circ}\times3^{\circ}$ image in the $170-231$ MHz sub-band, 
containing the Ophiuchus cluster. Yellow circles mark the $r=6^{\prime}.8$ 
regions used to estimate the contribution of the galactic foreground.}
\label{fig:gleambg}
\end{figure}
%%%%%%%%%%%%%%%%%%%%%%%%%%%%%%%%%%%%%%%%%%%%%%%%%%%%%%%%%%%%%%%%%%%%%

\section{Radio analysis}
\label{sec:flux}

\subsection{Flux density of the relic lobe}

We measured the flux density of the relic lobe in the GMRT, GLEAM, VLSSr 
and VLA images within a circular region of radius $r=6^{\prime}.8$, centered 
on RA$_{\rm J2000}$=17h12m50.2s and DEC$_{\rm J2000}$=$-23$d31m13s. This region 
corresponds to the X-ray cavity size inferred from the {\em XMM-Newton} 
image (Fig.~\ref{fig:xmm}) and is entirely filled by the relic lobe at 
240 MHz (Fig.~\ref{fig:images1}).
Table \ref{tab:flux} summarizes the relic lobe flux density at all
frequencies. Errors include the local image rms and flux calibration uncertainty 
(Eq.~\ref{eq:err}). Differences between the flux-density scales adopted for 
the GMRT and VLSSr (Scaife \& Heald 2012), GLEAM (Baars et al. 1977) and 
VLA data (Perley \& Butler 2014) are estimated to be at most $\sim 5\%$ 
(Perley \& Butler 2017).

For the VLSSr, we corrected the flux measured on the image for the clean
bias and adopted a flux uncertainty of 20\%, as appropriate for
extended sources detected at a very low signal-to-noise ratio (Lane et al.
2014).  This high uncertainty also accounts for the source location 
at the edge of the VLSSr mosaic, as fluxes can be affected by uncertainties 
in the primary-beam model at large distances from the pointing center.

For GLEAM, a systematic flux uncertainty of $8\%$ is expected for sources in
the same declination range as the Ophiuchus cluster (Hurley-Walker et al.
2017). Due to the very low Galactic latitude of Ophiuchus, the GLEAM 
fluxes reported in Table \ref{tab:flux} have been corrected for contamination 
due to Galactic diffuse synchrotron emission, to which GLEAM is highly sensitive
thanks to the very short $uv$ spacings of the MWA (e.g., Su et al.
2017). The GLEAM images of the region containing the Ophichus cluster exhibit 
in fact large-scale structures of Galactic emission, which affect 
flux density measurements. Fig.~\ref{fig:gleambg} shows a
$3^{\circ}\times3^{\circ}$ region, centered on Ophiuchus, of the 170-231 MHz
GLEAM image, where large variations of foreground emission are
visible across the image. In particular, the Ophiuchus cluster appears to be
located at the boundary of a bright region of Galactic emission. To estimate
the contribution of this Galactic foreground, we have extracted the flux
density within a set of 8 circular source-free regions of the same size as
the region used to measured the flux density of the relic lobe. As shown
in Fig.~\ref{fig:gleambg}, the regions were carefully placed around the
cluster to sample a wide range of surface brightness variations. The
following average values were obtained for each of the 4 GLEAM sub-band
images: $-2.41$ Jy, $-2.44$ Jy, $-0.89$ Jy and $+1.01$ Jy, going from the
lowest to the highest-frequency sub-band (since the large-scale Galactic
structure is not deconvolved in the GLEAM images, the Galactic contribution
can be positive or negative, Hurley-Walker et al. 2017).  Finally, these
values were used to correct the flux densities measured for the relic lobe
in the corresponding sub-band images.

%%%%%%%%%%%%%%%%%%%%%%%%%%%%%%%%%%%%%%%%%%%%%%%%%%%%%%%%%%%%%%%%%%%%%
\begin{table}
\caption{Flux density of the relic lobe}
\label{tab:flux}
\begin{center}
\begin{tabular}{lcc}
\hline\noalign{\smallskip}
\hline\noalign{\smallskip}
 & Frequency & Flux density \\
 &  (MHz)    & (Jy)       \\
\noalign{\smallskip}
\hline\noalign{\smallskip} 
VLSSr &  74   & $18.2\pm3.6$     \\
GLEAM &  88   & $11.6\pm1.8$  \\
GLEAM & 119   & $5.0\pm0.7$      \\
GMRT & 153    & $1.7\pm0.3$  \\
GLEAM & 154   & $2.1\pm0.5$      \\
GLEAM & 200   & $1.5\pm0.3$    \\
GMRT & 240 & $0.78\pm0.01$  \\
VLA & 1477 & $0.012\pm0.001$  \\
\hline{\smallskip}
\end{tabular}
\end{center}
Notes. -- At all frequencies, the flux densities were measured within 
the circular region shown in Fig.~\ref{fig:images1}, with $r=6^{\prime}.8$ and
centered on RA$_{\rm J2000}$=17h12m50.2s and DEC$_{\rm
  J2000}$=$-23$d31m13s. The VLSSr value has been corrected for clean bias and 
the GLEAM flux densities have been corrected for contamination due to diffuse 
Galactic synchrotron emission (see text). The flux densities from the GMRT 
and VLSSr images are on the Scaife \& Heald (2012) scale, the GLEAM fluxes are 
on the Baars et al. (1977) scale, and the 1477 MHz flux is on the Perley \& Butler 
(2014) scale. Differences between these scales are estimated to be at most
$\sim 5\%$ (Perley \& Butler 2017).
\end{table}
%%%%%%%%%%%%%%%%%%%%%%%%%%%%%%%%%%%%%%%%%%%%%%%%%%%%%%%%%%%%%%%%%%%%%

\subsection{Radio spectrum of the relic lobe}

\begin{figure}
\centering
\epsscale{1.1}
\includegraphics[width=9cm]{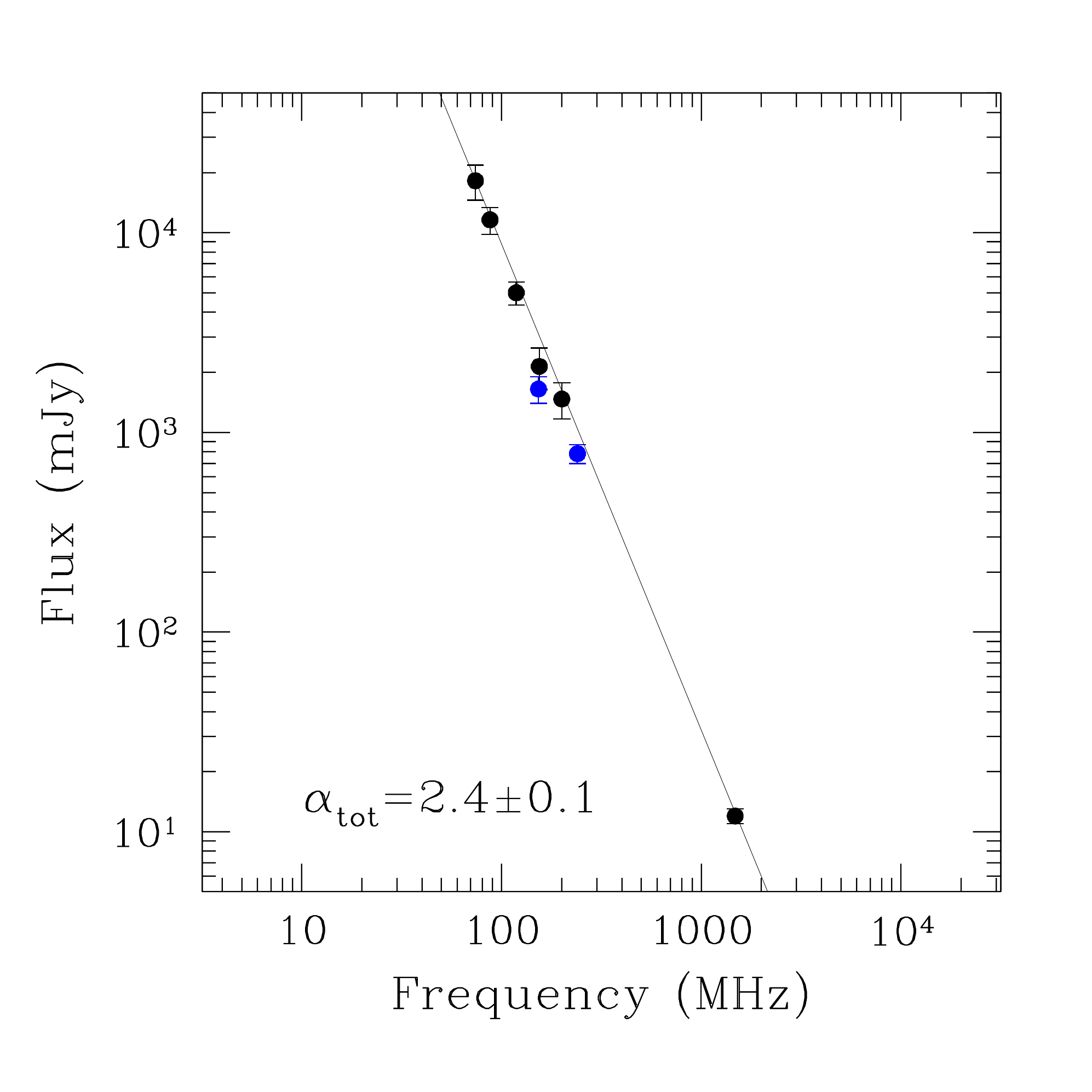}
\smallskip
\caption{Integrated radio spectrum of the relic lobe between 74 MHz and 
1477 MHz. The GMRT data points are shown in blue. 
The solid line is a power law with slope $\alpha_{\rm tot}=2.4\pm0.1$.}
\label{fig:sp}
\end{figure}

Using the flux density measurements in Table \ref{tab:flux}, we derived the
integrated radio spectrum of the relic lobe between 74 MHz and 1477 MHz,
shown in Figure \ref{fig:sp}. The spectrum is very steep and well described
by a power law with a spectral index $\alpha_{\rm tot}=2.4\pm0.1$. 
There is no indication of spectral curvature over the interval of frequency 
covered by our data. If we use only the low-frequency data points ($74-240$ MHz), 
we obtain a slope $\alpha_{\rm low}=2.7\pm0.2$, which is consistent within the 
errors with $\alpha_{\rm tot}$. Similarly, if we compute the spectral index 
from the GLEAM points alone, the resulting slope is $2.5\pm0.3$. The two GMRT 
measurements (blue points) give a flatter spectrum ($\alpha=1.7\pm0.4$). However, 
the 153 MHz flux is likely underestimated due to the missing structure in the 
relic lobe at 153 MHz compared to the 240 MHz image (Figs.~\ref{fig:images1} 
and ~\ref{fig:images2}). The 153 MHz flux density is also lower than the GLEAM flux 
density at a similar frequency of 154 MHz, though the two values are  still 
consistent within the error bars (Table \ref{tab:flux}).

%%%%%%%%%%%%%%%%%%%%%%%%%%%%%%%%%%%%%%%%%%%%%%%%%%%%%%%%%%%%%%%%%%%%%
%%%%%%%%%%%%%%%%%%%%%%%%%%%%%%%%%%%%%%%%%%%%%%%%%%%%%%%%%%%%%%%%%%%%%

\subsection{Radio spectrum of the minihalo}

M10 derived the integrated radio spectrum of the central minihalo and 
measured a possible steepening of the spectral index from 
$\alpha=1.4\pm0.3$ in the 153-240 MHz interval to $\alpha=1.60\pm0.05$ 
in the 240-1477 MHz range. However, the flux density values used by M10 
include part of the newly-discovered relic lobe. To investigate how this 
may impact the behavior of the minihalo spectrum, we rederived the minihalo 
flux densities after exclusion of the emission contributed by the relic lobe.
The faint emission associated with the minihalo is detected in both our 
GMRT images at 153 MHz and 240 MHz (Figs.~1 and 2). The brightest region 
of the minihalo is also detected in the GLEAM images. However, due to lower 
angular resolution, it is not possible to separate well the minihalo from 
the central radio galaxy and bright extended source north of the cluster center.
G19 measured a minihalo flux density of $62\pm9$ mJy at 1477 MHz, after 
subtraction of the contribution from discrete radio sources and relic lobe 
(see their Appendix A). Following G19, we measure a source-- and 
lobe--subtracted flux density of $1224\pm203$ mJy at 153 MHz and $692\pm70$ 
mJy at 240 MHz for the minihalo. The resulting total spectral index is 
$\alpha_{\rm tot,\, mh} = 1.32 \pm 0.10$ between 153 MHz and 1477 MHz. 
Contrary to the M10 finding of a possible steepening, the low-- and 
high--frequency indices are found to be consistent within the errors 
($\alpha_{\rm 153-240 \,MHz}=1.26 \pm 0.43$ and 
$\alpha_{\rm 240-1477 \, MHz}=1.33 \pm 0.10$), indicating that the minihalo
spectrum can be described by a single power law, at least up to 1477 MHz. 
High-sensitivity observations above this frequency are necessary to 
better constrain the overall shape of the minihalo spectrum and 
investigate the presence of a high-frequency steepening, as possibly seen 
in few other minihalos (e.g., Giacintucci et al. 2014).

\section{X-ray observations}
\label{sec:xray}

\subsection{Chandra}

The Ophiuchus cluster was observed by {\em Chandra}\/ ACIS-I in 2014 for a
total exposure of 230 ks (ObsIDs 16142, 16143, 16464, 16626, 16627, 16633,
16634, 16635, 16645). Results from this dataset were first presented by W16.
We extracted an image in the 0.5--4 keV band from these data using the
standard procedure (e.g., Wang et al.\ 2016). The image is shown in
  Figs.\ 1 and in Fig.\ \ref{fig:chandra} with different colors to
emphasize the curious concave X-ray brightness edge reported by W16; see
also Fig.\ 1 in W16. This rather sharp edge looks like a boundary of a
cavity in the X-ray gas in projection, although the elongation to the south
(``southern extension'' in Fig.\ 1) appears to be a separate hydrodynamic
disturbance and possibly even a small subcluster, as suggested by W16. We
point out in Sect.~\ref{sec:disc} that the cavity and the hydrodynamic
disturbance are not mutually exclusive.

The higher-resolution \gmrt\ 240 MHz contours are overlaid on the \chandra\ 
image in Fig.\ \ref{fig:chandra}{\em b}, which shows a striking spatial
coincidence between the inner edge of the relic lobe (the ``ridge'') and the
X-ray edge that is the boundary of the putative X-ray cavity.

\subsection{XMM-Newton}

The Ophiuchus cluster was observed with {\em XMM-Newton} for 37 ks in 2017
(ObsID 0505150101; Nevalainen et al.\ 2009). Several offset observations
were performed as well, which we do not use here. We extracted images from
this dataset using
the Extended Source Analysis Software ({\em XMM}-ESAS)%
\footnote{http://heasarc.gsfc.nasa.gov/docs/xmm/xmmhp\_xmmesas.html}
package part of SAS version 11.0.0. We used the standard data cleaning and
image technique (Snowden et al.\ 2008). The exposure-corrected and
background-subtracted 0.4--7.2 keV image combining the EPIC MOS and pn
detectors is shown in Fig.\ \ref{fig:xmm}. It shows the same concave
brightness edge (green arc) and allows us to trace it to somewhat larger
scales. The green circle in panel (d) is selected to trace this edge; its
radius is 230 kpc, which is somewhat bigger than in W16. This difference is
not meaningful as it is an extrapolation from a small sector, but our larger
radius also happens to encompass the giant low-frequency lobe that is seen
in GLEAM (Fig.\ \ref{fig:xmm}{\em d}) and GMRT lower-resolution (Fig.\ 
\ref{fig:xmm}{\em c}) images.

\subsection{ROSAT}
\label{sec:rosat}

It is interesting to check the larger-scale structure of the cluster, in
order to look for any ghost cavities on the opposite side. The \xmm\ offsets
do not cover the outskirts of the Ophiuchus cluster contiguously, so we have
checked the archives of other X-ray observatories looking for a better
large-scale image. \rosat\ PSPC and \asca\ have an interesting wide
coverage; we extracted a \rosat\ image because it has a better angular
resolution. We used an archival 0.5--2 keV image from a 4 ks \rosat\ PSPC
pointing toward the cluster center. It is shown in Fig.\ \ref{fig:rosat}
with contours from the lowest-band GLEAM image that shows the fossil lobe.
The X-ray statistics is clearly not sufficient to attempt a search for a
counter-cavity. However, the X-ray image shows that the overall cluster gas
distribution is elongated in the direction of the lobe, suggesting that the
radio and X-ray emission are physically related.

%%%%%%%%%%%%%%%%%%%%%%%%%%%%%%%%%%%%%%%%%%%%%%%%%%%%%%%%%%%%%%%%%%%%%
\begin{figure*}
\centering
\epsscale{1.1}
\includegraphics[width=8.5cm]{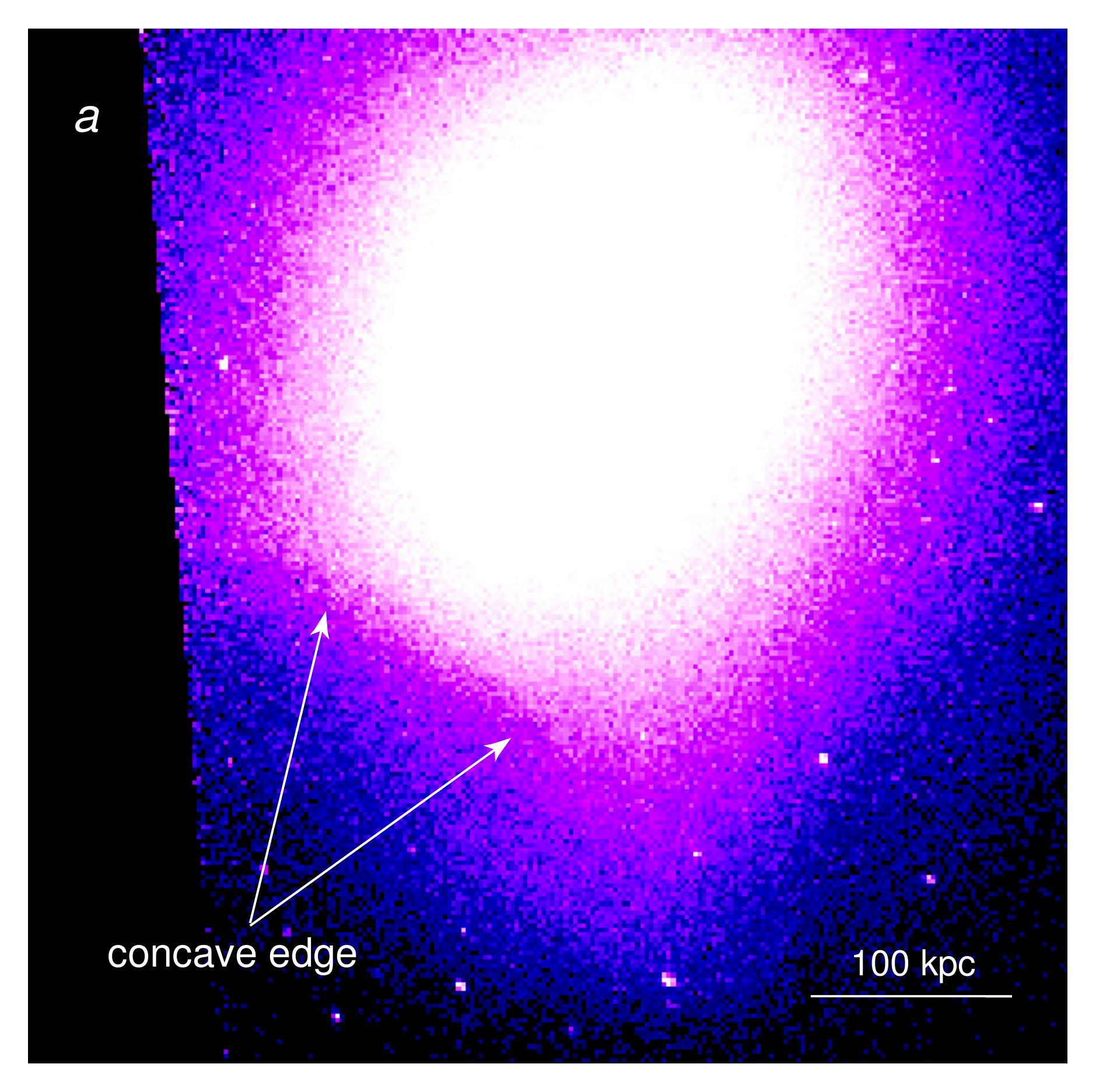}
\includegraphics[width=8.5cm]{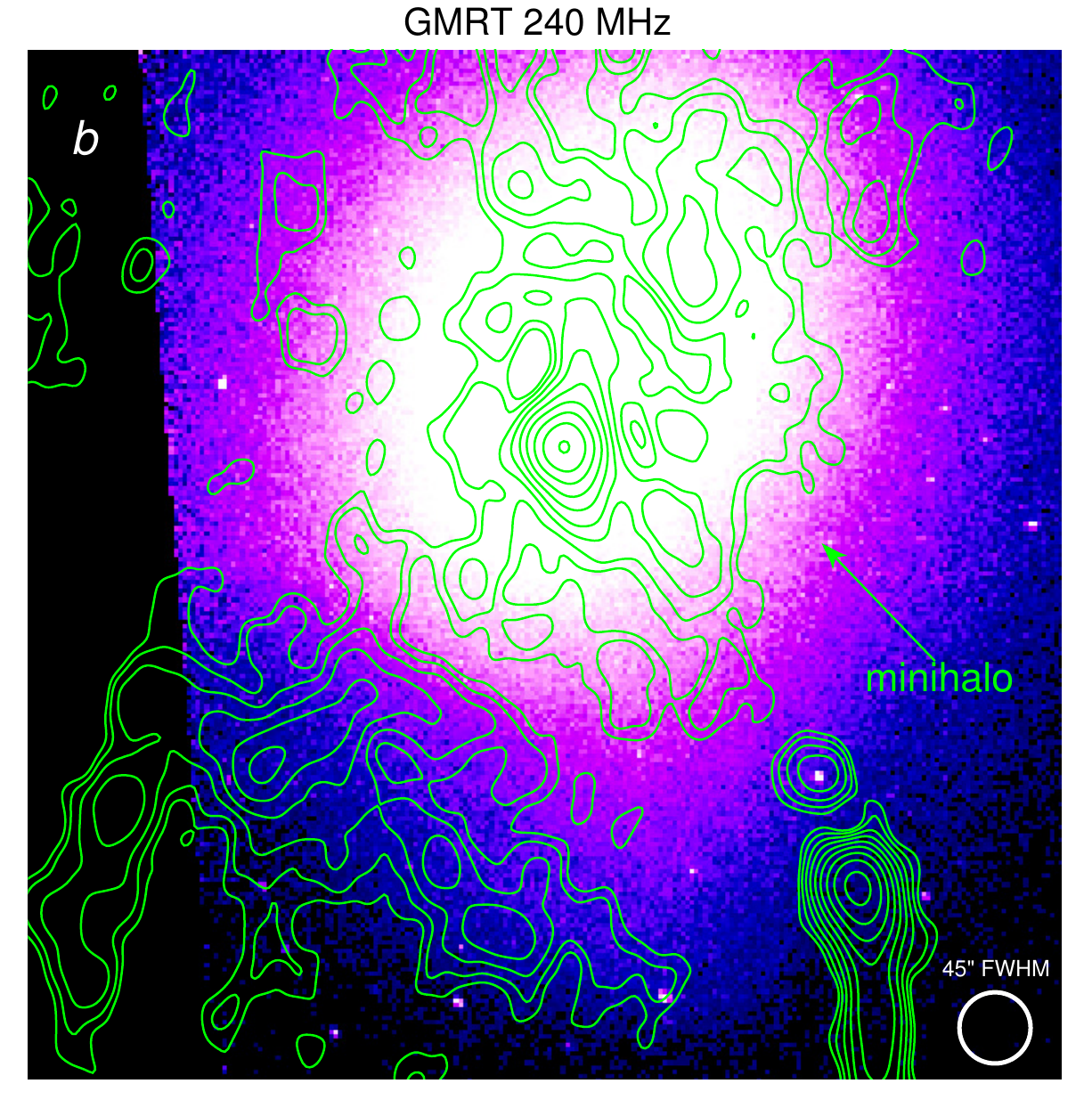}
\caption{\chandra\ X-ray image of the Ophiuchus cluster in the 0.5--4 keV
  band, binned to $4^{\prime\prime}$ pixels. (a) The concave edge, first
  reported by W16, is shown by arrows. (b) The GMRT 240 MHz image from
  Fig.~\ref{fig:images1}{\em a}\/ is overlaid as green contours.  The
  cluster core contains a radio minihalo. The extended source southeast of
  the core traces the X-ray edge.}
\label{fig:chandra}
\end{figure*}
%%%%%%%%%%%%%%%%%%%%%%%%%%%%%%%%%%%%%%%%%%%%%%%%%%%%%%%%%%%%%%%%%%%%%

%%%%%%%%%%%%%%%%%%%%%%%%%%%%%%%%%%%%%%%%%%%%%%%%%%%%%%%%%%%%%%%%%%%%%
\begin{figure*}
\centering
\epsscale{1.1}
\includegraphics[width=8.5cm]{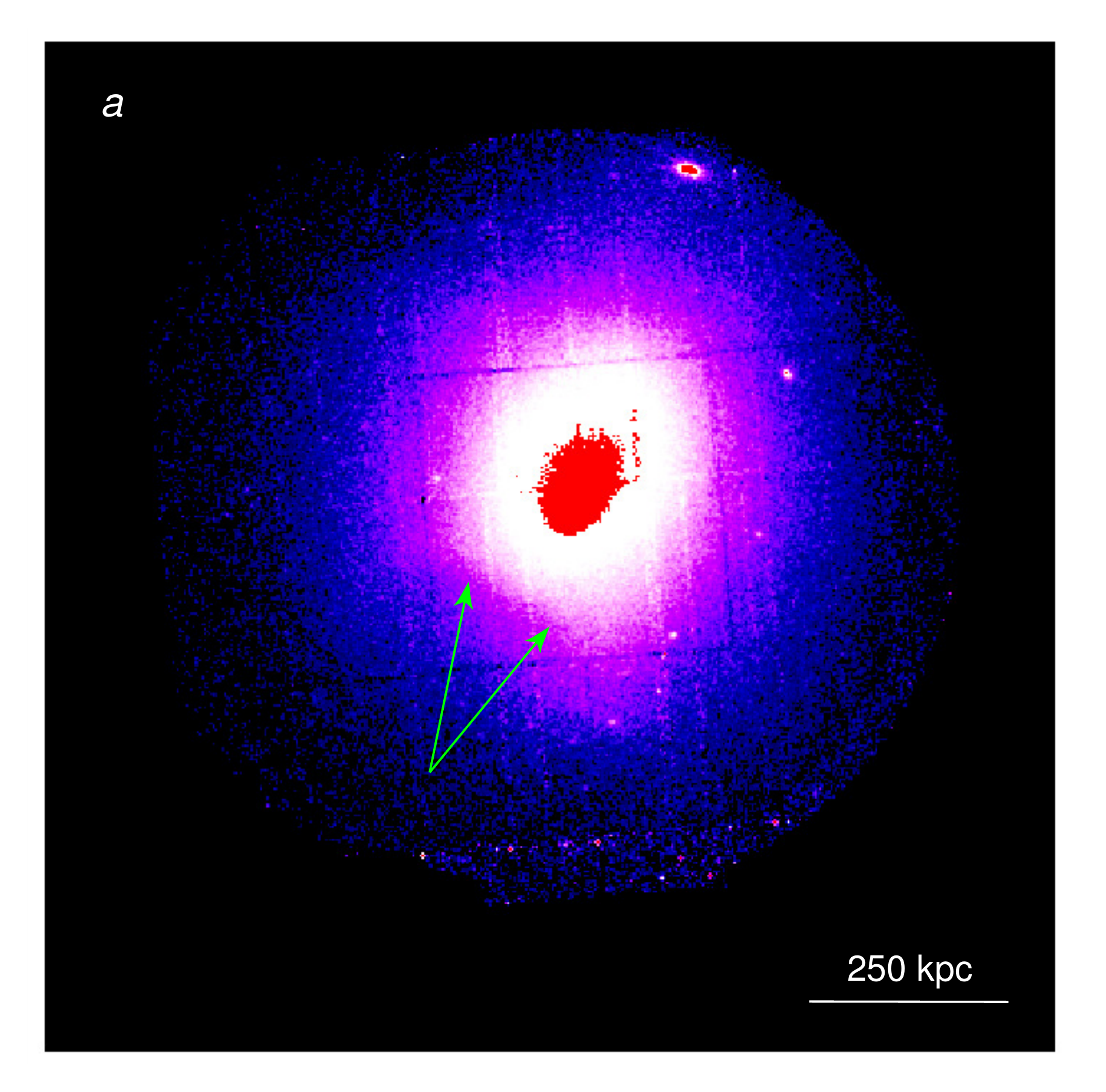}
\includegraphics[width=8.5cm]{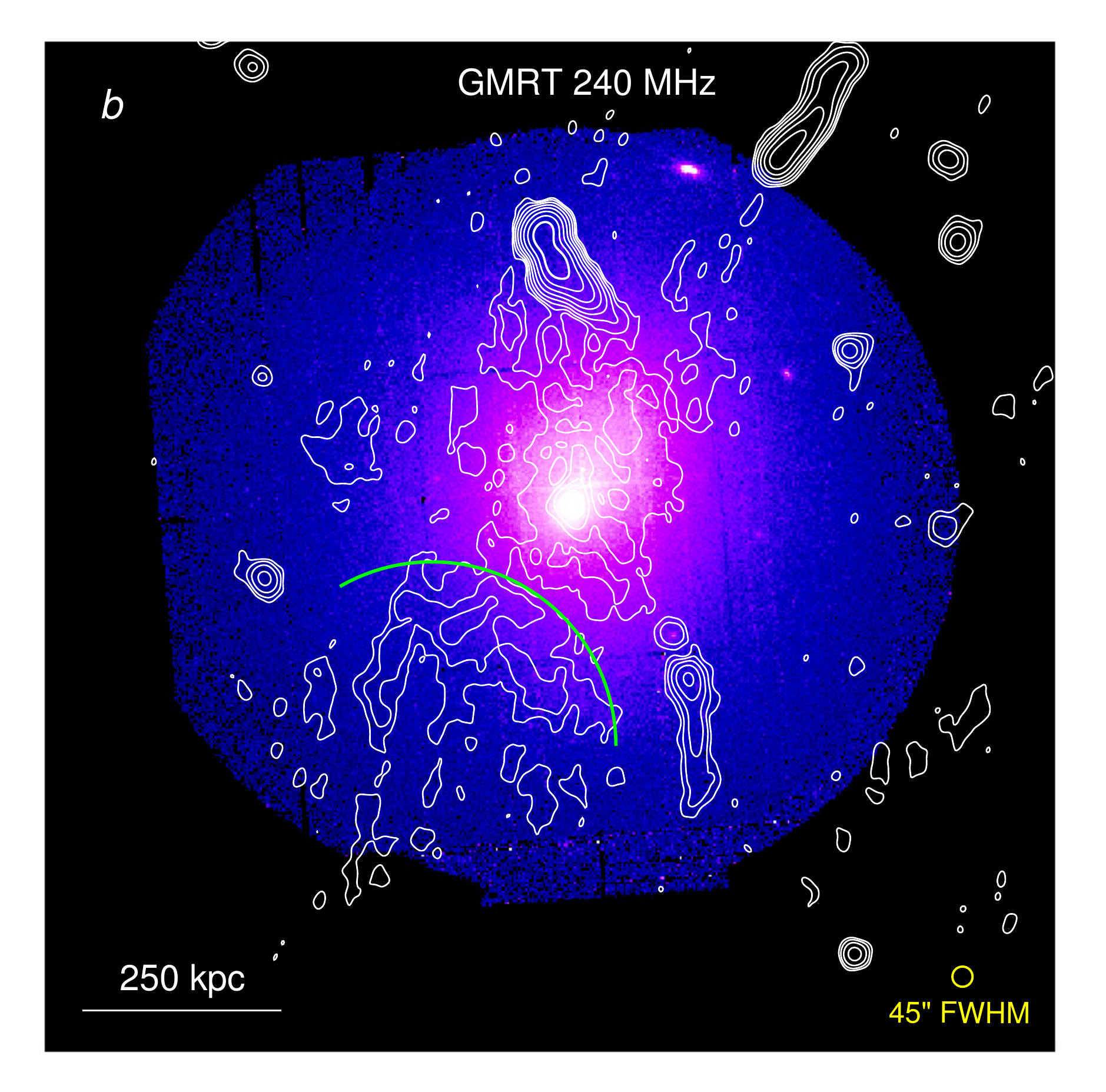}
\includegraphics[width=8.5cm]{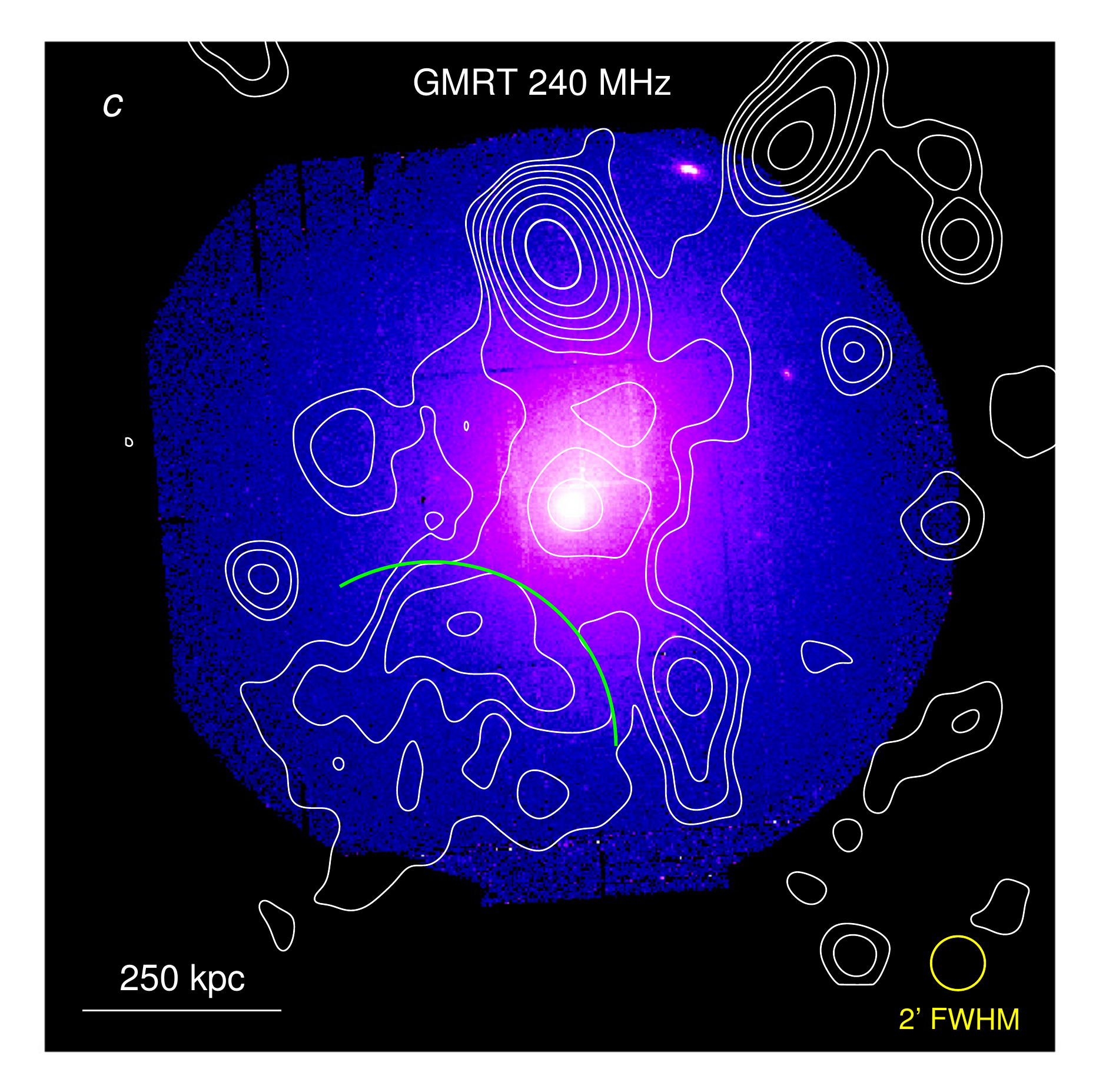}
\includegraphics[width=8.5cm]{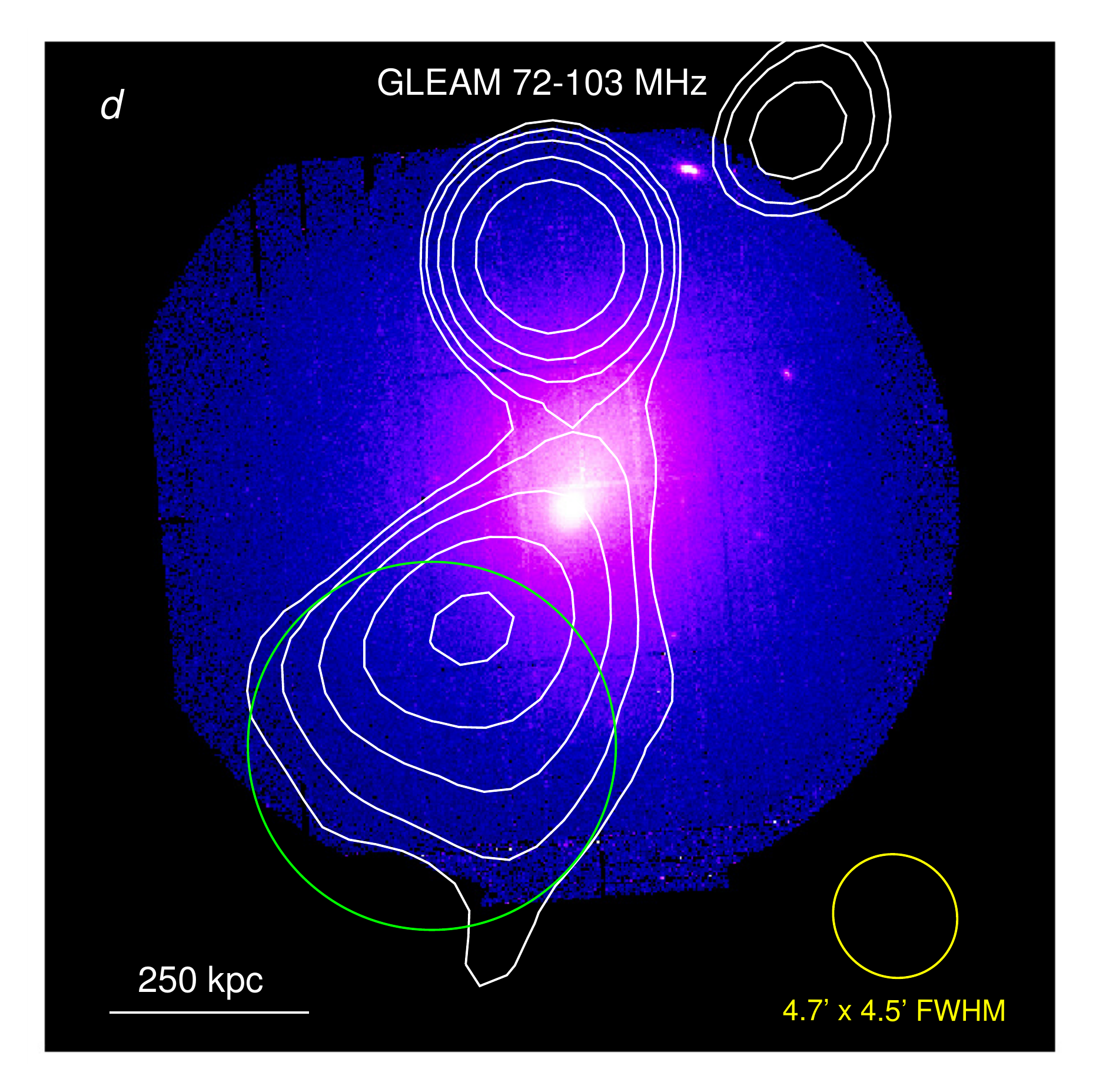}
\smallskip
\caption{(a) {\em XMM-Newton}\/ X-ray image of the Ophiuchus center in the
  0.4--7.2 keV band, combining pn and MOS detectors, binned to 5\as\ pixels.
  The concave edge is shown with arrows (as in Fig.~\ref{fig:chandra}).
  (b,c,d) The same {\em XMM-Newton}\/ image using a different color scale,
  with the radio contours overlaid from Figs.\ \ref{fig:images1} and
  ~\ref{fig:gleam}{\em a}, spaced by a factor 2 starting from $+3\sigma$.
  Radio beams are shown in yellow.  The diffuse radio emission southeast of
  the cool core fills the apparent X-ray cavity and at the lowest frequency
  becomes a dominant diffuse source. The green circle in (d) encompasses
  this radio lobe and at the same time traces the X-ray edge in the \xmm\ 
  and \chandra\ images. The green arcs in (b,c) are a portion of this circle
  that traces the edge.}
%: (b) GMRT, 240 MHz, 45\as\ beam,
%  from Fig.\ \ref{fig:images1}{\em a}; (c) GMRT, 240 MHz, 2\am\ beam,
%  from Fig.\ \ref{fig:images1}{\em b}; (d) GLEAM 72--103 MHZ, $4'\!.5$ beam,
%  from Fig.~\ref{fig:gleam}{\em a}. {\bf ADD beam circles and labels
%    GMRT/MWA, freq.\ to images, then can omit beams from caption.} 
%All radio contours are spaced by a factor 2 starting from $+3\sigma$.
%, which
%  corresponds to (b) $1.65$ mJy beam$^{-1}$, (c) $9.9$ mJy beam$^{-1}$ and
%  (d) $297$ mJy beam$^{-1}$. 

%
\label{fig:xmm}
\end{figure*}
%%%%%%%%%%%%%%%%%%%%%%%%%%%%%%%%%%%%%%%%%%%%%%%%%%%%%%%%%%%%%%%%%%%%%

%%%%%%%%%%%%%%%%%%%%%%%%%%%%%%%%%%%%%%%%%%%%%%%%%%%%%%%%%%%%%%%%%%%%%
\begin{figure*}
\centering
\epsscale{1.1}
\includegraphics[width=15cm]{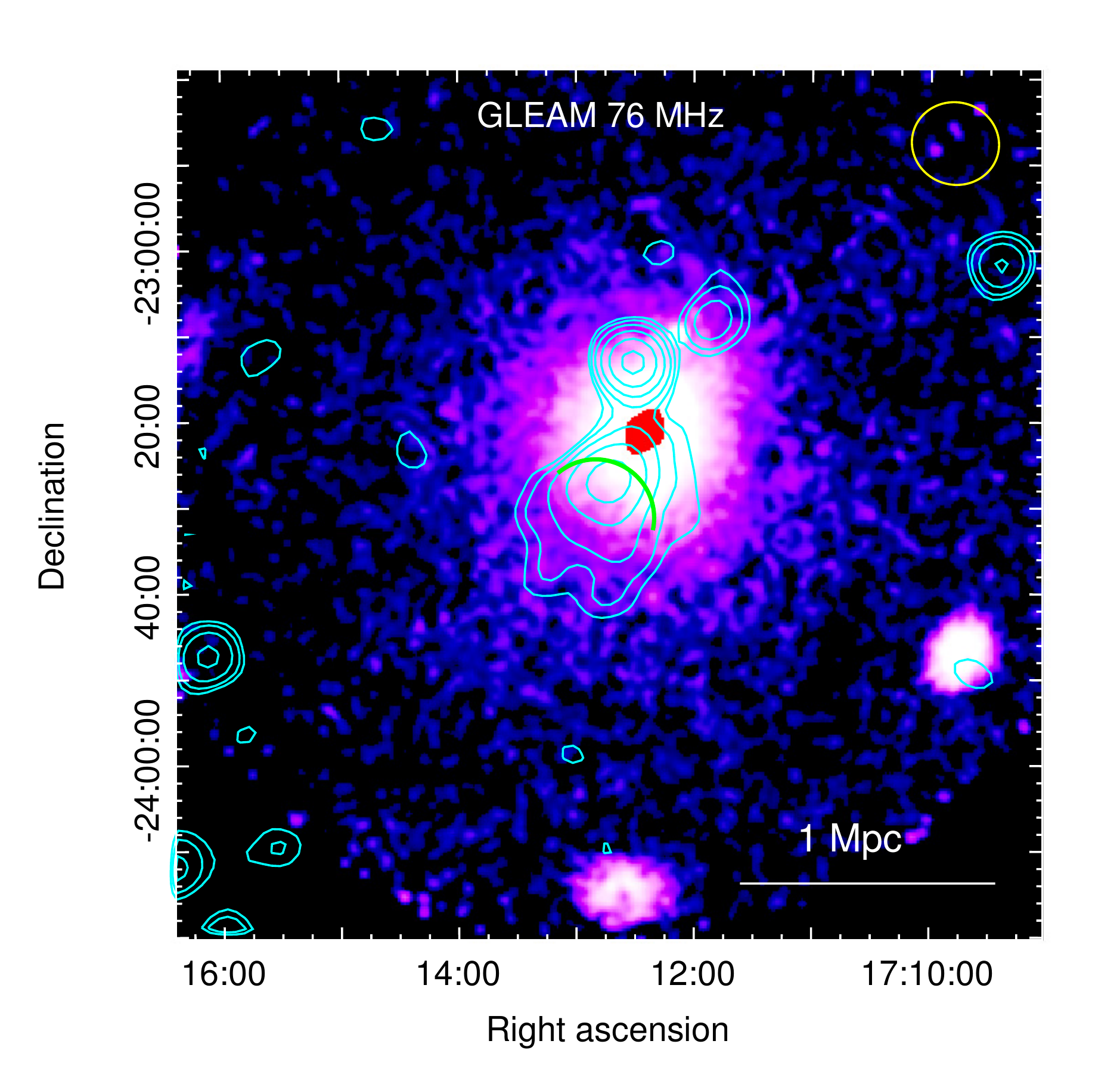}
\caption{\mwa\ GLEAM contours in the 72-80 MHz band overlaid on the 
\rosat\ PSPC X-ray image that shows the cluster structure on larger scales. 
The cluster shows elongation on the side of the radio lobe. The green arc 
traces the X-ray edge seen in the \chandra\ and \xmm\ images. (The two X-ray blobs at bottom and lower-right are point sources affected 
by the degraded angular resolution at the edges 
of the \rosat\ field of view.) Radio contours are spaced by 
a factor of 2 starting from 400 mJy beam$^{-1}$. 
The radio beam (yellow ellipse) is $5^{\prime}.1\times4^{\prime}.8$, 
in p.a. 15$^{\circ}$.}
\label{fig:rosat}
\end{figure*}
%%%%%%%%%%%%%%%%%%%%%%%%%%%%%%%%%%%%%%%%%%%%%%%%%%%%%%%%%%%%%%%%%%%%%

%%%%%%%%%%%%%%%%%%%%%%%%%%%%%%%%%%%%%%%%%%%%%%%%%%%%%%%%%%%%%%%%%%%%%
\begin{figure*}
\centering
\epsscale{1.1}
\includegraphics[width=8.5cm]{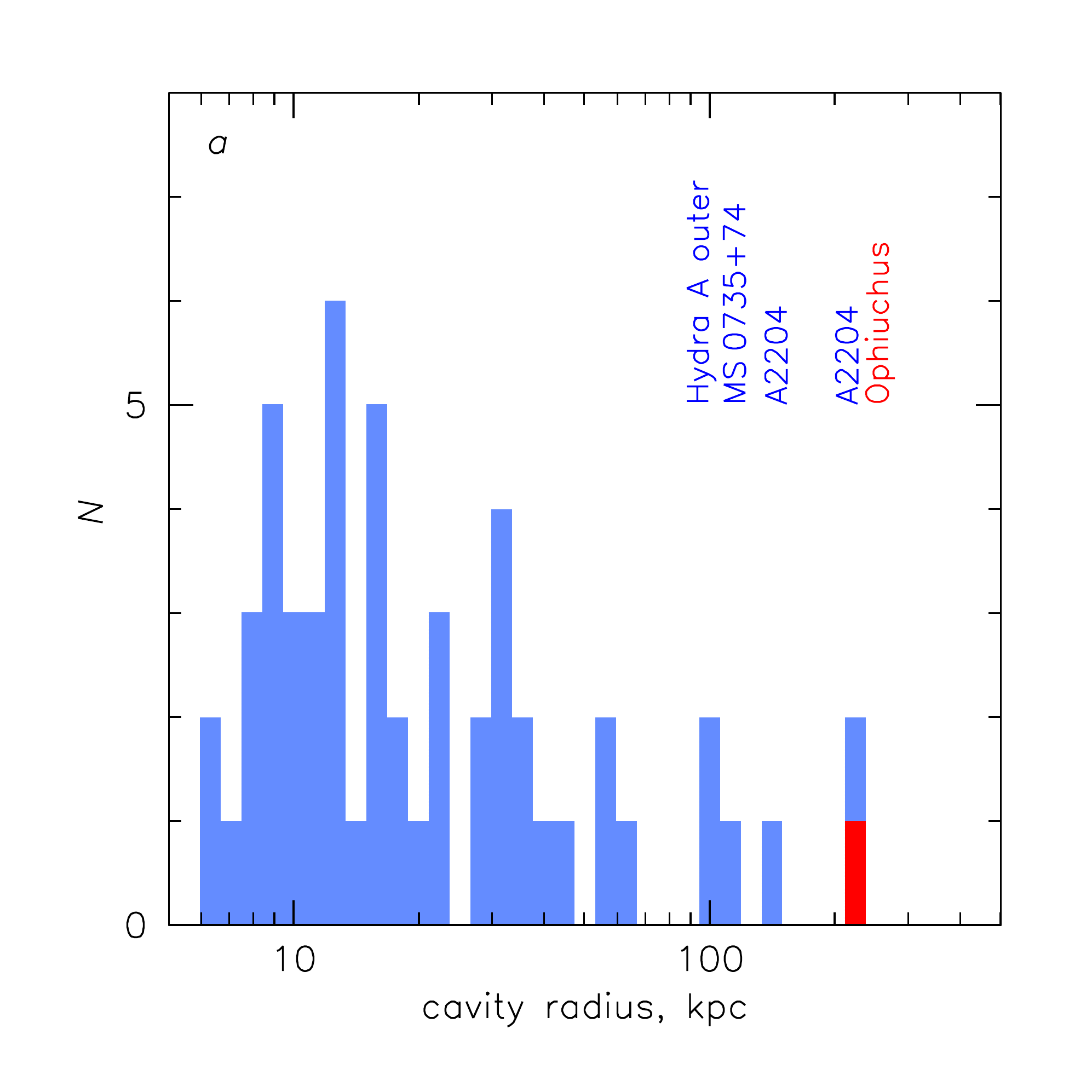}
\includegraphics[width=8.5cm]{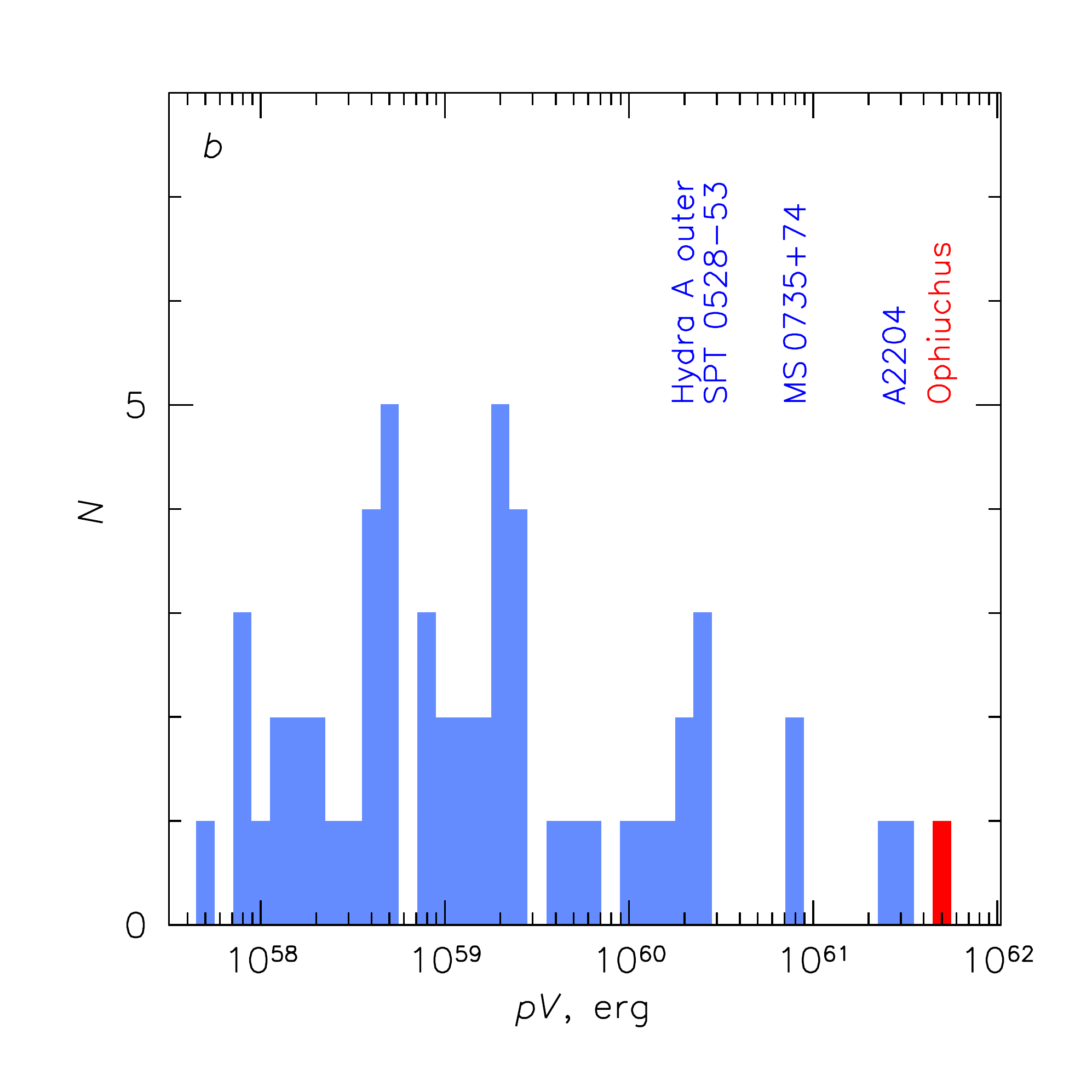}
\caption{The distribution of known AGN cavities in galaxy clusters: (a)
  by their size and (b) by their energy ($pV$\/ for each cavity). The
  biggest cavities are labeled. Data from Rafferty et al.\ (2006)
  supplemented by the Hydra A outer cavities from Wise et al.\ (2007),
the recently-discovered cavities in SPT-CLJ0528-5300 from Calzadilla et al.\ (2019) 
and the possible ghost cavities in A\,2204 from Sanders et al.\ 
  (2009). The Ophiuchus cavity $pV$\/ estimate is from W16, while the size 
is from this work (we find a somewhat larger size than W16, thus 
the $pV$\/ value may be an underestimate). The X-ray based cavity sizes 
are uncertain and the $pV$\/ estimates even more so, but the Ophiuchus cavity is certainly an
  enormous outlier.}
\label{fig:cav}
\end{figure*}
%%%%%%%%%%%%%%%%%%%%%%%%%%%%%%%%%%%%%%%%%%%%%%%%%%%%%%%%%%%%%%%%%%%%%

\section{Discussion}
\label{sec:disc}

\subsection{Fossil of an enormous AGN outburst?}

The comparison of the radio images and the high-resolution \chandra\ and
\xmm\ X-ray images (Figs.\ \ref{fig:chandra},\ref{fig:xmm}) shows a large
low-frequency radio lobe that fits into the X-ray edge like hand in glove.
The spatial coincidence is especially striking between the X-ray edge and
the higher-resolution \gmrt\ image (Fig.\ \ref{fig:chandra}b). Neither
\chandra\ nor \xmm\ existing images have the spatial coverage or sensitivity
to trace the outer boundaries of the ICM cavity (which would look like a
very low-contrast X-ray brightness edge), but the curvature of the visible
edge implies that the GLEAM and \gmrt\ radio lobe would fill this putative
cavity (Fig.\ \ref{fig:xmm}) as seen in many clusters with AGN lobes, such
as Perseus.

What makes this cavity unique is its size and the energy required to create
it. The latter has been estimated by W16 to be $pV\sim 5\times 10^{61}$ erg,
where $p$\/ comes from the pressure of the ICM around it and $V$\/
from the \chandra\ X-ray edge curvature. This value is of course only an
order-of-magnitude estimate, as it involves several assumptions, such as
where the cavity is located on the line of sight (which should not be too
far from the cluster sky plane because it is still visible in the image) and
the ICM density and temperature profiles at the relevant 3D radii. With that
caveat in mind, it is interesting to compare the cavity $pV$\/ and size
with the AGN-blown cavities in other clusters. We show such a comparison in
Fig.\ \ref{fig:cav}, using data for individual cavities (as opposed to their
pairs --- since we see only one in Ophiuchus) from Rafferty et al.\ (2006),
adding the outer cavities in Hydra A (Wise et al.\ 2007), the recently-discovered 
cavities in SPT-CLJ0528-5300 (Calzadilla et al.\ 2019) and a pair of giant
cavities in A2204 (Sanders et al.\ 2009; these are seen in the X-ray but not
in the radio, so their nature is still ambiguous). The cavity in Ophiuchus
would require an AGN outburst that is an extreme outlier in terms of its
total energy -- orders of magnitude more energetic than those typically seen
in clusters, and 5 times more energetic than the previous record holder,
MS\,0735+74.

\subsection{The age of the lobe}

If the lobe has started its life as a buoyant bubble injected near the
cluster center by the Ophiuchus central galaxy, it would take it at least
240 Myr to rise to the current radius moving with the sound speed, which is
an upper limit on the velocity, so the actual age would be greater than that.

Another method often used to estimate the age of radio galaxy lobes is to
look for a break in their radio spectrum and model it. The highest-energy
relativistic electrons cool first and the electron power spectrum (and the
spectrum of their synchrotron emission) develops an exponential cutoff,
which moves to lower energies (and lower radio frequencies) as time passes
(e.g., Komissarov and Gubanov 1994). While the radio spectrum of the Ophiuchus 
cavity is very steep, it is
described well by an unbroken power law across our range of frequencies from
74 MHz to 1477 MHz, and we do not see any high-frequency cutoff (\S\ref{sec:flux}). 
We cannot simply assume that the cutoff is outside our frequency band, 
because for an aged source of such an enormous linear size, the total synchrotron spectrum 
is likely to be the sum of a patchwork of local spectra of different shapes, 
which could add up to a powerlaw. A large and old radio lobe is likely to 
contain very nonuniform, filamentary magnetic fields, which would produce different 
synchrotron spectra and different local synchrotron cooling times even for 
a single underlying electron power spectrum (e.g., Tregillis, Jones, Ryu 2004). 
The resulting total radio synchrotron spectrum thus cannot be usefully related 
to the spectrum of the emitting electrons (e.g., Eilek \& Arendt 1996; Katz, Stone
\& Rudnick 1997).

In this regard, it may be useful to look at a spectral index map with
interesting spatial resolution, which would at least remove the spectral
averaging over the face of the lobe. We have recently obtained new, deeper
and more sensitive observations of this source with the upgraded GMRT (uGMRT) 
in Band 2 (130--260 MHz) and Band 3 (250--500 MHz) and we will present 
the analysis of these new observations in a forthcoming paper.

\subsection{The origin of the cavity}

The spatial coincidence of the X-ray cavity (or at least its inner edge) and
the giant, steep-spectrum radio lobe is striking and strongly suggests that
we have uncovered a fossil of an enormous AGN outburst. However, there are
several puzzles in this scenario. First, the central AGN, the supposed
culprit, is rather faint and does not show any jets (W16), nor are there 
any other more powerful AGNs in the vicinity of the relic lobe. However,
the central AGN could have been much more active in the past. The X-ray 
image shows that sloshing has currently displaced the peak of the cluster 
cool gas away from the cD galaxy, something that is rarely seen in clusters 
and that can temporarily starve the AGN of its accretion fuel (Hamer et al. 
2012, W16). The gas sloshing seen in the X-ray image may even have been 
set off by the powerful outburst that has created the giant lobe.

A more difficult puzzle is that we see only one lobe, whereas AGN jets
usually come in symmetric pairs and produce a pair of radio lobes in
the ICM on two sides of the AGN. Given the likely advanced age of our lobe,
we speculate that the counterpart may have propagated into a less-dense ICM
on the other side of the cluster and completely faded away. Furthermore, the
lobes move together with the ICM in which they are embedded, which breaks
the symmetry of the pair (e.g., the outer lobes in Hydra A, Nulsen et al.\ 
2005). Strong cluster-wide bulk gas motions, such as those generated by a
merger, may have moved the counterpart lobe somewhere in the outskirts,
where it faded. An exotic alternative is that the radio galaxy that gave rise to this
bubble is not the cD galaxy, but some other galaxy perhaps in the cluster
outskirts; the counterpart lobe would then fade quickly in the low-density
gas far from the cluster center.

Another serious problem was posed by W16 --- how did the cool core survive
such an energetic event? There are clusters whose cores appear to have been
torn to pieces by AGN outbursts (e.g., A2390, Vikhlinin et al.\ 2005). 
The Ophiuchus cool core has a steeper radial density and temperature gradient 
than most cool cores (W16), which may make it more stable against a disruption 
that comes from the side rather than from the nucleus. One can imagine long 
jets from the central AGN piercing the cool core and depositing energy and 
blowing bubbles right outside the core. On smaller linear scales, this is 
observed in some cluster galaxies with dense gaseous coronae 
(e.g., Sun et al. 2007, Sanders et al. 2014) or in remnant compact cool cores 
(e.g., Cheung et al. 2019), which are pierced by jets emitted by the AGN 
without disrupting the corona/compact core, and creating bubbles right outside.
%We note that in this scenario, the age estimate that we made above
%based on the off-center distance is not meaningful.

The Ophiuchus core does exhibit stronger sloshing motions than most cool
cores. It may have been triggered by an asymmetric pair of expanding bubbles
deposited outside the core, which may have given the core a push from the
side while avoiding destroying the core. It would be interesting to
explore such a possibility using hydrodynamic simulations.

On larger linear scales, the \rosat\ image (\S\ref{sec:rosat}) exhibits a
curious elongation in the exact direction of the giant lobe. There are two
interesting possibilities here. One is the analogy with Hydra A, where a
pair of giant radio bubbles is ensconced in an X-ray cocoon (see Fig.\ 2 in
Nulsen et al.\ 2005), whose outer edge is a weak shock front (Simionescu et
al.\ 2009). Another possibility is cluster-wide sloshing, such as recently
found in Perseus (Simionescu et al.\ 2012) and A2142 (Rossetti et al.\ 
2013), both of which exhibit cold fronts $\sim 1$ Mpc from the center. If
the enormous AGN outburst managed to move the cool core and make it slosh,
it might induce cluster-wide sloshing as well --- its mechanical energy
approaches that of minor mergers.

We note here that the hydrodynamic instability of a sloshing front, proposed
by W16 to explain the concave brightness edge, and our radio bubble
explanation, are not mutually exclusive. The ``subcluster,'' or the surface
brightness excess south of the core at one end of the concave feature (Fig.\ 
1b in W16 and ``southern extension'' in our Fig.\ 1), exhibits a
morphology that makes it likely to be a splash of cool gas from the core
generated by the expansion of the bubble and the resulting major
hydrodynamic disturbance of the cluster.

\section{Conclusions}

The ``gentle'' AGN feedback is the current preferred solution for the
cluster cooling flow problem. In this scenario, the mechanical energy of multiple 
outbursts from the central AGN provides gentle, steady,
self-regulated heating of the cool gas in the core that exactly compensates
for fast X-ray radiative cooling of that gas (e.g., 
McNamara \& Nulsen 2012; Fabian 2012, e.g.; Voit et al. 2015; Gaspari et al.\ 2017).  
Ophiuchus offers a dramatic example that is far outside this
scenario. We found evidence of an extraordinarily powerful past AGN outburst
that has deposited its energy outside the cool core, without completely
disrupting the core. The evidence comes in the form of a giant radio lobe
with a very steep spectrum, discovered by \mwa\ and confirmed by \gmrt,
which coincides spatially with an apparent cavity in the cluster gas. 

This enormous fossil radio source may be an early example of a new
class of sources to be uncovered by the ongoing highly sensitive
low-frequency surveys of galaxy clusters. Their counterpart cavities
in the X-ray gas outside the cluster bright central regions will have
very low brightness contrast. To detect them in the X-ray, it would
require a combination of high angular resolution and large
collecting area to obtain the sufficient photon statistics. For example,
{\em Athena}'s imaging instrument (Barcons et al.\, 2017)
%(https://www.the-athena-x-ray-observatory.eu/) 
and the {\em AXIS}\/ concept (Mushotzky et al.\, 2019) will be able 
to detect such features.
\\
\\
{\it Acknowledgements.}

We thank the anonymous referee. This scientific work makes use of the 
Murchison Radio-astronomy Observatory, operated by CSIRO. We 
acknowledge the Wajarri Yamatji people as the traditional owners of 
the observatory site. Support for the operation of the MWA is provided 
by the Australian Government (NCRIS), under a contract 
to Curtin University administered by Astronomy Australia Limited.
We acknowledge the Pawsey Supercomputing Centre which is supported by the 
Western Australian and Australian Governments. GMRT is run by the National Centre 
for Astrophysics of the Tata Institute of Fundamental Research. The National Radio 
Astronomy Observatory is a facility of the National Science Foundation operated 
under cooperative agreement by Associated Universities, Inc. Basic research in radio 
astronomy at the Naval Research Laboratory is supported by 6.1 Base funding.
Q. H. S. Wang acknowledges the support of NASA via {\em Chandra} grant AR5-16013X.

{}

\end{document}